\documentclass[useAMS,usenatbib]{mn2e}
\usepackage{lscape}
\usepackage{natbib}
\usepackage[pdftex]{graphicx}

\DeclareMathVersion{bold}

\def\lsim{\mathrel{\rlap{\lower 3pt\hbox{$\sim$}}\raise 2.0pt\hbox{$<$}}}
\def\gsim{\mathrel{\rlap{\lower 3pt\hbox{$\sim$}} \raise 2.0pt\hbox{$>$}}}

\title[Blazars in H-ATLAS]{Mining the \textsl{Herschel}\thanks{{\it Herschel} is an ESA space observatory with science instruments provided by European-led Principal Investigator consortia and with important participation from NASA.}-ATLAS: submillimeter-selected blazars in equatorial fields}
\date{\vspace{-10ex}}
 \author[M. L\'opez-Caniego et al.]{\parbox[t]{\textwidth}
{M. L\'opez-Caniego$^{1}$\thanks{E-mail: caniego@ifca.unican.es}, J. Gonz\'alez-Nuevo$^{1}$, M. Massardi$^{2}$, L. Bonavera$^{1}$, D. Herranz$^{1}$, M. Negrello$^{5}$, G. De Zotti$^{5,6}$, F. J. Carrera$^{1}$, L. Danese$^{6}$, S. Fleuren$^{3}$, M. Hardcastle$^{13}$, , M. J. Jarvis$^{13,19}$, H.-R. Kl\"ockner$^{15,16}$, T. Mauch$^{13}$ , P. Procopio$^{4}$, S. Righini$^{2}$, W. Sutherland$^{3}$, R. Auld$^{7}$, M. Baes$^{14}$, S. Buttiglione$^{5}$, C. J. R. Clark$^{7}$,  A. Cooray$^{12}$, A. Dariush$^{8}$, L. Dunne$^{17}$, S. Dye$^{9}$, S. Eales$^{7}$ ,R. Hopwood$^{8,18}$, C. Hoyos$^{9}$, E. Ibar$^{10}$,  R. J. Ivison$^{10,11}$, S. Maddox$^{17}$, E. Valiante$^{7}$ }\\
\vspace*{4pt} \\	
$^{1}$Instituto de F\'\i{sica} de Cantabria (CSIC-UC), Avda. los Castros s/n, 39005 Santander, Spain\\
$^{2}$INAF- Istituto di Radioastronomia, Via P. Gobetti n. 101, 40129 Bologna, Italy\\
$^{3}$School of Physics and Astronomy, Queen Mary University of London, Mile End Road, London, E1 4NS, UK\\
$^{4}$School of Physics, David Caro Building, Corner of Tin Alley \& Swanston St, The University of Melbourne, Parkville, VIC 3010, Australia\\
$^{5}$INAF-Osservatorio Astronomico di Padova, Vicolo dell'Osservatorio 5, I-35122 Padova, Italy\\
$^{6}$Astrophysics Sector, SISSA, Via Bonomea 265, 34136 Trieste, Italy\\
$^{7}$School of Physics and Astronomy, Cardiff University, The Parade, Cardiff, CF24 3AA, UK\\
$^{8}$Department of Physics, Imperial College London, South Kensington Campus, London SW7 2AZ,UK\\
$^{9}$School of Physics and Astronomy, University of Nottingham, University Park, Nottingham NG7 2RD, UK\\
$^{10}$UK Astronomy Technology Centre, Royal Observatory, Blackford Hill, Edinburgh EH9 3HJ, UK \\
$^{11}$Institute for Astronomy, University of Edinburgh, Royal Observatory, Edinburgh EH9 3HJ, UK \\
$^{12}$Department of Physics and Astronomy, University of California, Irvine, CA 92697, USA \\
$^{13}$Centre for Astrophysics Research, STRI, University of Hertfordshire, Hatfield AL10 9AB \\
$^{14}$Sterrenkundig Observatorium, Universiteit Gent, Krijgslaan 281 S9, B-9000 Gent, Belgium \\
$^{15}$University of Oxford, Denys Wilkinson Building, Oxford OX1 3RH \\
$^{16}$Max-Planck-Institut f\"ur Radioastronomie, Auf dem H\"uegel 69, 53121 Bonn, Germany \\
$^{17}$Department of Physics and Astronomy, University of Canterbury, Private Bag 4800, Christchurch, 8140, New Zealand\\
$^{18}$Department of Physical Sciences, The Open University, Milton Keynes MK7 6AA, UK\\
$^{19}$Physics Department, University of the Western Cape, Private Bag X17, Bellville 7535, South Africa}
\begin{document}

\maketitle

\vspace*{200pt}

\begin{abstract}
The \textsl{Herschel} Astrophysical Terahertz Large Area Survey (H-ATLAS) provides an unprecedented opportunity to search for blazars at sub-mm wavelengths. We cross-matched the FIRST radio source catalogue with the 11655 sources brighter than 35 mJy at $500\,\mu$m in the $\sim 135$ square degrees of the sky covered by the H-ATLAS equatorial fields at 9\,h and 15\,h, plus half of the field at 12\,h.  We found that 379 of the H-ATLAS sources have a FIRST counterpart within 10 arcsec, including 8 catalogued blazars (plus one known blazar that was found at the edge of one the H-ATLAS maps). To search for additional blazar candidates we have devised new diagnostic diagrams and found that known blazars occupy a region of the $\log(S_{500\mu\rm m}/S_{350\mu\rm m})$ vs. $\log(S_{500\mu\rm m}/S_{1.4\rm GHz})$ plane separated from that of the other sub-mm sources with radio counterparts. Using this diagnostic we have selected 12 further candidates that turn out to be scattered in the $(r-z)$ vs. $(u-r)$ plane or in the WISE colour-colour diagram proposed by Massaro et al. (2012), where known blazars are concentrated in well defined strips. This suggests that the majority of them either are not blazars or have spectral energy distributions contaminated by their host galaxies. A significant fraction of true blazars are found to be hosted by star-forming galaxies. This finding, supported by an analysis of blazars detected in  {\it Planck} 545 and 857 GHz bands, is at odds with the notion that blazar hosts are passive ellipticals and indicates that the sub-mm selection is providing a novel prospect on blazar properties. Based on an inspection of the available photometric data, including the WISE all-sky survey, the unpublished VIKING survey and new radio observations, we tentatively estimate that there are 11 blazars with synchrotron flux density $S_{500\mu\rm m}> 35\,$mJy over the considered area. This result already allows us to constrain blazar evolution models.
\end{abstract}

\section{Introduction} \label{sec:intro}
Blazars are a subclass of active galactic nuclei (AGNs) characterized by non-thermal continuum emission from radio to  $\gamma$-rays. They have extreme properties: they are strongly variable across the full electromagnetic spectrum, reach very high observed luminosities and are frequently strongly polarized. Their  spectral energy distribution (SED) is characterized by two broad peaks in \textbf{$\nu L_\nu$}. The first peak occurs at a frequency $\nu_{p}^s$ varying from $\sim 10^{12}\,$Hz to $\sim 10^{19}\,$Hz ($\sim 300 \mu m$ to $\sim 3\times10^{-5} \mu m$)  \citep{nieppola06} and is attributed to Doppler boosted synchrotron emission from a highly relativistic jet pointing towards the observer. The continuum spectrum is ``flat'' (spectral index $\alpha \gsim -0.5$, $S_\nu \propto \nu^\alpha$) from radio to sub-mm or even shorter wavelengths. The second peak, attributed to inverse Compton scattering, occurs at $\gamma$-ray energies.

However, the non-thermal emission not always outshine other components, like emission from the host galaxy, from the  accretion disk and from the circum-nuclear torus, to the point of making such components undetectable. Blazar hosts are generally found to be early-type galaxies \citep{Kotilainen1998,ODowdUrry2005,Kotilainen2007,LeonTavares2011} and may therefore contaminate the non-thermal spectrum primarily at optical/near-IR wavelengths. However, we should also take into account the possibility that at least some of them are endowed with star formation activity associated with thermal dust emission at far-IR/sub-mm wavelengths. The direct thermal emission from the accretion disk may show up as a  relatively narrow `bump' in the optical/UV \citep{DermerSchlickeiser1993,GhiselliniTavecchio2009} while the thermal emission from the dusty circum-nuclear torus may be detectable in the mid-IR  \citep{Perlman2008}.

The optical spectra of blazars show a striking dichotomy, leading to the recognition of two sub-classes: BL Lacertae objects (BL Lacs) with an almost featureless spectrum, and Flat-Spectrum Radio Quasars (FSRQs) with strong, broad emission lines. \cite{padovani95} classified the BL Lacs on the basis of the synchrotron peak frequency, $\nu_{p}^s$,  into low, intermediate and high frequency peaked. This classification was extended by \cite{abdo10}  to all blazars: low synchrotron peak (LSP:  $\nu_{p}^s<10^{14}\,$Hz), intermediate synchrotron peak (ISP: $10^{14}< \nu_{p}^s < 10^{15}\,$Hz) and high synchrotron peak  (HSP: $\nu_{p}^s>10^{15}\,$Hz) objects. The distribution of $\nu_{p}^s$ is  bimodal, with only a minor fraction of objects peaking at intermediate frequencies. The bimodality however may be, at least partly, due to the blazar selection which, so far, was mostly done either in the radio or in the X-ray band. The former selection favours LSPs, the latter HSPs.

The {\it Herschel} selection, at intermediate frequencies, should help us to pick up ISPs, thus providing constraints on the abundance of these objects and indications on whether or not there is a continuity in the blazar population, from LSPs to HSPs. Moreover, a full frequency coverage is crucial to disentangle the different, non-thermal and thermal, contributions to the blazar SEDs, as briefly summarized above. The \textsl{Herschel} Astrophysical Terahertz Large Area Survey \citep[H-ATLAS;][]{eales10}, the largest area survey carried out by the \textit{Herschel} space observatory \citep{pilbratt10} covering $\sim 550\,\hbox{deg}^2$ with PACS \citep{poglitsch11} and SPIRE \citep{griffin10} instruments between 100 and $500\,\mu$m, provides an unprecedented opportunity to obtain flux density limited blazar samples at sub-mm wavelengths. As for the non-thermal emission, the \textsl{Herschel} data cover a particularly interesting frequency range close to the synchrotron peak of the most luminous LSPs \citep{fossati98}, where information is currently scanty \citep{padovani06}. Since $\nu_p^s$ carries information on the Lorentz factor $\gamma_p$ of emitting electrons, on the Doppler factor $\delta$ and on the magnetic field strength $B$ ($\nu_p\propto \gamma_p^2\, \delta\, B$), its value provides crucially important information on key physical parameters. Moreover, the sub-mm selection is especially well suited to look for signatures of thermal dust emission from the host galaxy, thus providing a new perspective on blazar hosts and allowing us to test the early-type host paradigm.

Blazars are however a tiny fraction of H-ATLAS sources: their estimated surface density at the H-ATLAS $5\,\sigma$ detection limits is $\sim 0.1$ percent or less of the surface density of dusty galaxies. Singling them out is not easy and appropriate diagnostic tools must be devised. An investigation of the H-ATLAS Science Demonstration Phase (SDP) field \citep{gnuevo10} uncovered two blazars brighter than the average $5\sigma$ detection limit at $500\,\mu$m, $S_{500\mu\rm m}\simeq 44\,$mJy \citep{rigby11}, over an area of $14.4\,\hbox{deg}^2$. In this paper we extend the analysis to an area almost an order of magnitude larger. This allows us to build the first statistically meaningful sample of sub-mm selected blazars.

The layout of this paper is the following. In Section\,\ref{sec:candidates} we describe the selection of candidate blazars, present the new diagnostic diagrams we have devised, and report on the follow-up observations aimed at assessing the nature of the selected candidates. In Section\,\ref{sec:SEDs} we briefly discuss their SEDs, built by combining the H-ATLAS photometry with literature and unpublished VIKING data. In Section\,\ref{sec:counts} we discuss constraints on sub-mm blazar counts and implications for evolutionary models. Finally, Section\,\ref{sec:conclusions} summarizes our main results.

We adopt a flat cosmology with $\Omega_{\Lambda}=0.734$, $\Omega_{m}=0.266$, and $h=0.71$ as derived from WMAP 7-yr data \citep{Larson2011}.

\begin{figure*}
\begin{center}
\includegraphics[width=0.8\textwidth]{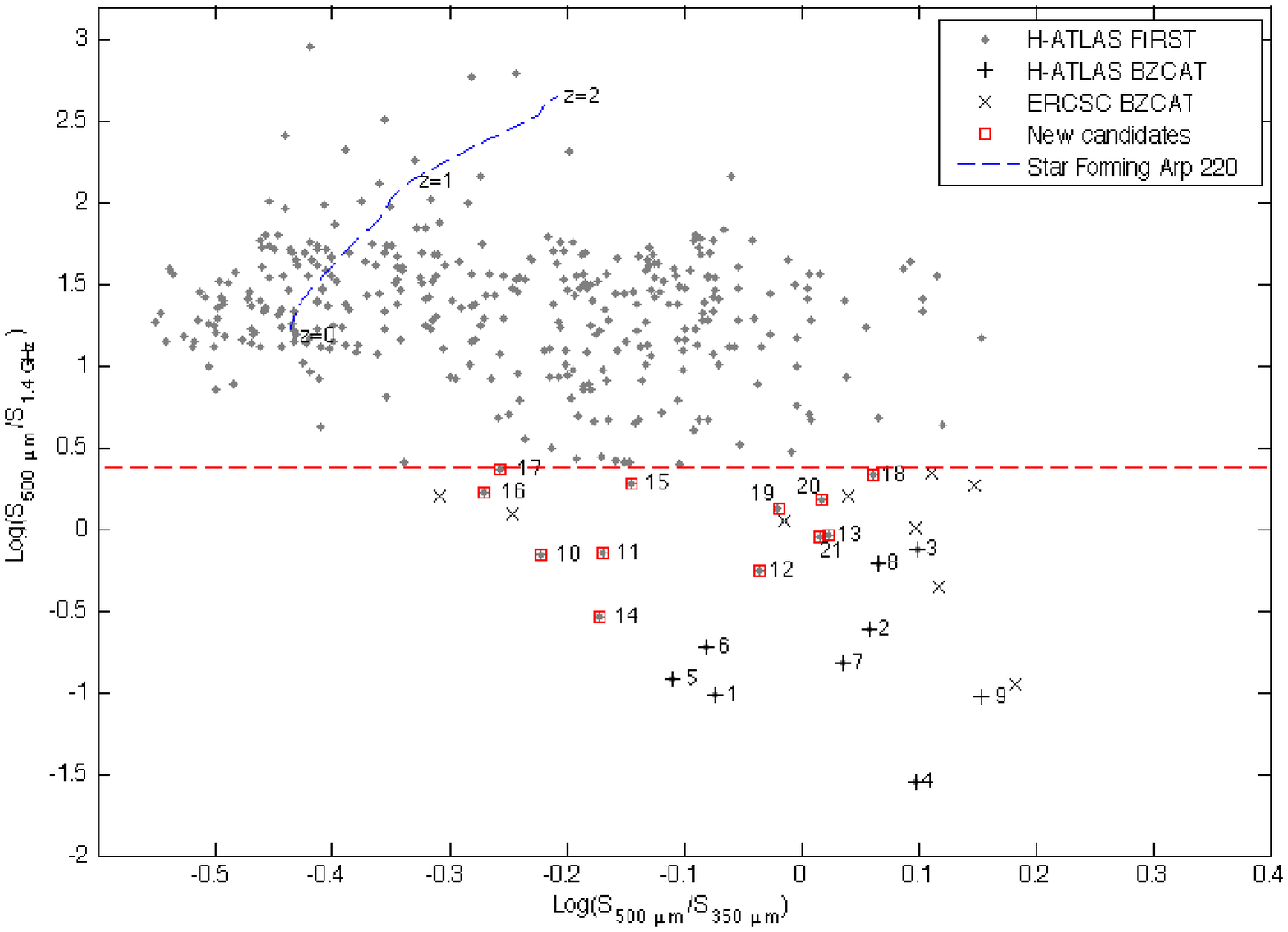}
\includegraphics[width=0.8\textwidth]{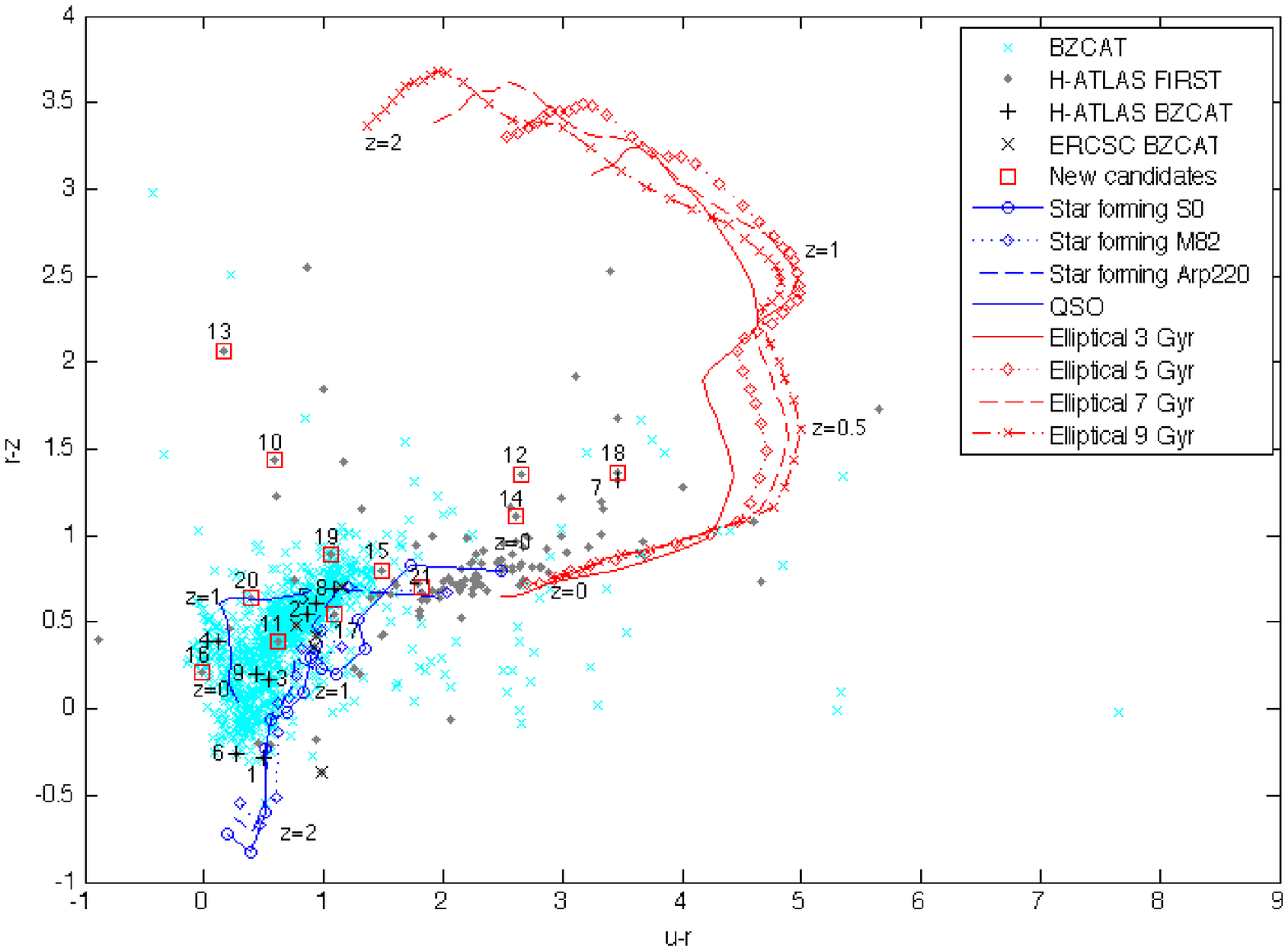}
\end{center}
\caption{Diagnostic colour-colour diagrams: $S_{500\mu{\rm m}}/S_{1.4\rm GHz}$ vs. $S_{500\mu{\rm m}}/S_{350\mu\rm m}$ (upper panel) and $(r-z)$ vs. $(u-r)$ (lower panel). In both panels the grey dots denote the H-ATLAS sources with $S_{500\mu\rm m}\ge 35\,$ mJy and FIRST counterparts (only a subset of these sources have SDSS photometry and thus appear also in the lower panel), except for the 9 catalogued blazars that are identified by the larger black $+$ signs. The grey $+$ signs enclosed in red squares denote the new blazar candidates. Both catalogued and candidate H-ATLAS blazars are labelled with their ID numbers (see Tables~\protect\ref{tab:tabla_confirm8} and \protect\ref{tab:tabla_candidates}). The {\it Planck} ERCSC blazars detected both at 545 and at 857 GHz (see text) are represented by the black $\times$ signs. In the upper panel the thick dashed line shows the colour-colour track of a prototype star-forming galaxy, Arp~220. The errorbars in $log(S_{500\mu{\rm m}}/S_{1.4\rm GHz})$ for the catalogued and candidate blazars are $\leq 0.1$ and have not been plotted for the sake of clarity. In the lower panel, the light blue $\times$ signs correspond to BZCAT blazars with SDSS photometry. In this panel we also show, for comparison, the colour-colour tracks, as a function of redshift, of 4 passive elliptical templates of different ages (red lines) and of 3 star-forming galaxies (blue lines). The S0, M82, QSO and Arp220 tracks were computed using the SEDs tabulated in the SWIRE library \citep{polletta2007} available at http://www.iasf-milano.inaf.it/$\sim$polletta/templates/swire\_templates.html. The elliptical galaxy tracks were computed using GALSYTH, the web based interface for the GRASIL model \citep{silva1998} available at  http://galsynth.oapd.inaf.it/galsynth/. The colours corresponding to some redshift values are indicated.  }\label{fig:col_plot}
\end{figure*}

\section{Blazar candidates} \label{sec:candidates}

In this study we use the catalogue of SPIRE sources detected in the two H-ATLAS equatorial fields centred at about 9\,h and 15\,h, plus half of that at 12\,h, with areas of $53.25\,\hbox{deg}^2$, $53.93\,\hbox{deg}^2$, and $27.37\,\hbox{deg}^2$, respectively, for a total of $134.55\,\hbox{deg}^2$. In the Phase 1 data release [Hoyos et al. (in preparation), Valiante et al. (in preparation)] data for the full 12\,h field will be included. The catalogue  contains sources that have been detected in the noise-weighted PSF-filtered 250 $\mu$m map applying the MADX algorithm \citep{maddox12}. Then, at the position of the sources, flux densities are estimated at each of the SPIRE bands. Finally, only sources with a signal-to-noise $\geq 5$ at any of the SPIRE bands are included in the catalogue. The H-ATLAS maps and data reduction are discussed in \citet{pascale11} and \citet{ibar10}, the source catalogue creation is described in \citet{rigby11}. The SPIRE beams have FWHM of $18.1$, $24.8$ and $35.2$ arcsec for 250, 350 and $500\,\mu$m respectively \citep{Swinyard2010,rigby11}. The catalogue we have used contains 66535 unique objects (26369 in the 9h field, 12686 in the 12h field, and 27480 in the 15h field). Since blazars have red sub-mm colours they are more easily singled out at the longer wavelengths. We have therefore confined ourselves to sources with $S_{500\mu\rm m} \ge 35\,$mJy, {corresponding to $4\sigma$ detections} at this wavelength. There are 11655 sources satisfying this criterion.
\begin{figure}
\begin{center}
\includegraphics[width=0.5\textwidth]{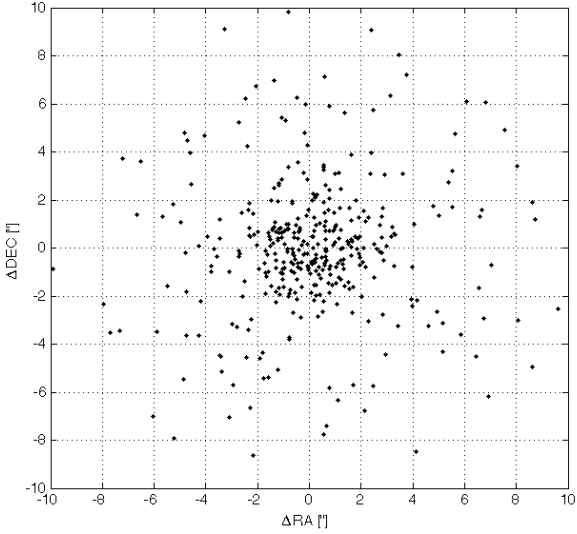}
\end{center}
\caption{Positional offset between the 379 H-ATLAS sources with $S_{500\mu\rm m} \ge 35\,$mJy and their FIRST counterpart within $10$ arcsec. }\label{fig:ra_dec_offset}
\end{figure}
Blazars are powerful radio sources. Thus the first step towards spotting them is to require that they are detected by radio surveys covering our fields. A cross correlation of these sources with the FIRST (Faint Images of the Radio Sky at Twenty-cm) survey catalogue \citep{becker95}  with a $10$ arcsec search radius yields 379 matches, out of the $\sim 11000$ FIRST sources present in the $\sim 135\,\hbox{deg}^2$ H-ATLAS equatorial fields.

The sample made of the 379 sources selected as described above ($S_{500\mu\rm m} \ge 35\,$mJy and FIRST counterpart within $10$ arcsec) will be referred to as the parent sample. The typical rms positional error for our sources is $2.5$ arcsec or less \citep{rigby11}\footnote{The quoted positional error
refers to $> 5\,\sigma\ 250\,\mu m$ sources as most (355) of the 379 sources
in the parent sample are.}, and the FIRST positions are accurate to better than $0.5$ arcsec for $S_{1.4\rm GHz}> 3\,$mJy, so that the rms offset between the radio and the sub-mm position is $\sigma \le 2.55$ arcsec (see Fig.~\ref{fig:ra_dec_offset}). In the case of a Gaussian distribution of positional errors the probability that a true counterpart has an apparent positional offset $\ge \Delta$ is
\begin{equation}\label{eq:prob}
p(>\Delta) = \exp[-0.5(\Delta/\sigma)^2].
\end{equation}
For $\Delta =10$ arcsec, $p(>\Delta)\le 4.6\times 10^{-4}$ and the number of true FIRST counterparts that we may have missed with our 11655 trials is $\simeq 5$,  i.e. $\simeq 1.5$ percent  of the parent sample.

The surface density of FIRST sources is $n_{\rm FIRST}\simeq 90\,\hbox{deg}^{-2}$. Therefore the probability that a FIRST source lies by chance within $\Delta =10$ arcsec from a given \textit{Herschel} source is  $\pi \Delta^2 n_{\rm FIRST}\simeq 2\times 10^{-3}$ and the expected number of chance associations is $\simeq 25$. However, as we will see in the following, from the distribution of $S_{500\mu\rm m}/S_{1.4\rm GHz}$ flux density ratios of known blazars we infer that genuine blazars must have $S_{500\mu\rm m}/S_{1.4\rm GHz}< 2.4$ or, in our case, $S_{1.4\rm GHz}>14\,$mJy. The surface density above this limit is $\simeq 13.5\,\hbox{deg}^{-2}$ and the number of chance associations decreases to $\simeq 4$.
\begin{figure}
\begin{center}
\includegraphics[width=0.47\textwidth]{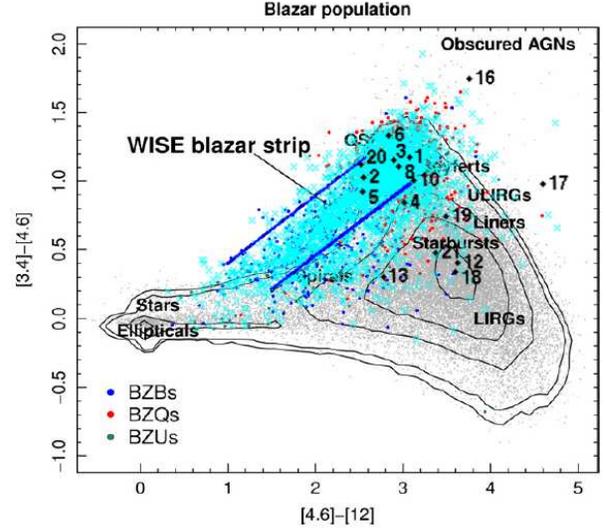}
\end{center}
\caption{Diagnostic diagram using WISE colours. BZCAT objects are shown in cyan and our candidate and confirmed blazars are labelled with their ID numbers (see Tables~\protect\ref{tab:tabla_confirm8} and \protect\ref{tab:tabla_candidates}). The other data points and the boundaries of the `WISE blazar strip' are from \protect\cite{Massaro2011}. BZB, BZQ, and BZU stand for BL Lac, flat-spectrum radio quasar, and blazar of uncertain type, respectively. The contours tracing the density of sources in colour bins with annotations showing the location of different classes of objects are from the WISE Explanatory Supplement  \protect\citep{Cutri2011}.}
\label{fig:WISE_diag}
\end{figure}

\subsection{Diagnostic diagrams}\label{sec:blazar_phot}

A characteristic property of blazars is their peculiar continuum spectrum. This property allows us to construct diagnostic diagrams useful to select the tiny fraction of blazar candidates out of the overwhelming number of dusty galaxies. The relatively flat non-thermal continuum of blazars from radio to sub-mm wavelengths implies that they have red sub-mm colours and sub-mm to radio flux density ratios substantially lower than those of dusty galaxies. Thus we expect that blazars occupy in, e.g.,  the $\log(S_{500\mu\rm m}/S_{350\mu\rm m})$ vs $\log(S_{500\mu\rm m}/S_{1.4\rm GHz})$ plane a region separated from that populated by sub-mm galaxies dominated by thermal dust emission.

This expectation is borne out by the upper panel of Fig.~\ref{fig:col_plot}. To locate the ``blazar region'' in this diagram we have looked for known blazars among the 379 sources in the parent sample by cross-matching them with the most recent version of the blazar catalogue BZCAT \citep{Massaro2011BZCAT}, again with a search radius of $10$ arcsec. We got 8 matches (see Table~\ref{tab:tabla_confirm8}). Since the BZCAT positions are mostly taken from radio catalogues and are generally better than $1$ arcsec, after eq.~(\ref{eq:prob}) the expected number of real matches that may have been missed with our 379 trials is $\le 0.4$. The probability that a BZCAT source lies by chance within $10$ arcsec from a given \textit{Herschel} source is difficult to assess accurately because of the inhomogeneous distribution of BZCAT sources. However the BZCAT catalogue contains $\simeq 3100$ sources over the whole sky so that their mean surface density is $n_{\rm BZCAT}< 0.1\,\hbox{deg}^{-2}$. If they were uniformly distributed the expected number of chance associations would be $<10^{-3}$. There is therefore a wide enough margin to confidently assume that all the associations found are real even though the distribution is not uniform. In particular, of the 8 matches found with a search radius of $10$ arcsec, the two with the largest angular distance to a FIRST source (objects $\# 1$ and $2$ of Table~\ref{tab:tabla_confirm8}, with a separation of $8.67$ and $9.72$ arcsec, respectively) have $S_{1.4\rm GHz} > 300$ mJy. Since the surface density of FIRST sources brighter than 300 mJy is $\sim 0.46\,\hbox{deg}^{-2}$, the probability of a chance association within $10$ arcsec is $\sim 1.1\times10^{-5}$ and the expected number of chance associations within $10$ arcsec is 0.13. On the other hand, after eq.~(\ref {eq:prob}) the probability that a true counterpart has a positional offset of 9.72 arcsec
or of 8.67 arcsec is $\simeq 7\times 10^{-4}$ or $\simeq 3\times 10^{-3}$, respectively. Thus the blazar classification of these 2 H-ATLAS sources must
be taken with caution. 

In the BZCAT catalogue there are an additional 15 blazars located in our fields. For only two of them could we detect a significant signal in the H-ATLAS noise-weighted beam convolved maps. One (BZBJ0849+0206) has a $500\,\mu$m flux density of 31.6 mJy ($\hbox{S/N}\simeq 3.6$) and is weaker at 250 and $350\,\mu$m ($\hbox{S/N}\simeq 2.1$ and 2.2, respectively). Since this object is fainter than the adopted flux-density limit ($S_{500\mu{\rm m}}=35\,$mJy) we have not considered it further. The other object  (BZQJ1404-0013) has $S_{500\mu{\rm m}}=45\,$mJy, i.e. is a $5\sigma$ detection at this wavelength, but is below the $5\sigma$ detection limit at 250 and $350\,\mu$m ($S_{250\mu{\rm m}}\simeq 18\pm 7\,$mJy and $S_{350\mu{\rm m}}\simeq 32\pm 8\,$mJy). It was probably missed because of its low $250\,\mu$m flux density \citep[candidate sources were identified as $>2.5\sigma$ peaks at $250\,\mu$m;][]{rigby11}; moreover, it lies close to the border of the field. We have added this source to our blazar sample. We have also searched the 2-yr {\it Fermi}-LAT catalogue\footnote{fermi.gsfc.nasa.gov/ssc/data/access/lat/2yr\_catalog/} \citep{2FGL} for counterparts of our sources that might be blazars not listed in the BZCAT. None was found: the only {\it Fermi}-LAT source located within our fields and not listed in the BZCAT is identified as a pulsar and is about $3$ arcmin away from the nearest source in the parent sample.

To increase the sample of known blazars with measured sub-mm colours we have cross-correlated with the BZCAT the full-sky {\it Planck} Early Release Compact Source Catalogue \citep[ERCSC;][]{ercsc}, which consists of sources with flux density measurements at both 545 and 857 GHz. We found 14 matches The additional data available in the NASA Extragalactic Database\footnote{http://ned.ipac.caltech.edu/} (NED) show that one of these sources, BZUJ1325-4301, %(BZCAT \#~1784)
is the well known extended radio galaxy Centaurus A, while the blazar classification of BZBJ1136+1601 %BZCAT \#~1417
is dubious and BZUJ2209-4710 (alias NGC~7213) %\#~2800
has sub-mm colours indicative of a strong contamination by thermal dust emission from the host galaxy. Moreover, the ERCSC flux densities of BZQJ1559+0304 and BZQJ1719+0817 %BZCAT \#~2223 and \#~2411
may be affected by source blending. The flux density ratios for these sources are therefore not representative of the blazar non-thermal continuum. We are then left with 9 blazars (see Table~\ref{tab:tabla_ercsc}), whose colours are also plotted in the upper panel of Fig.~\ref{fig:col_plot}. According to the technical specifications of their respective instruments, the {\it Planck}/HFI 857 GHz and the {\it Herschel}/SPIRE $350\,\mu$m channels have almost exactly the same central wavelength and roughly the same bandwidth. The comparison between the {\it Planck} 545 GHz and the SPIRE $500\,\mu$m flux densities requires a colour correction because the central frequency of the $500\,\mu$m channel is somewhat higher ($\simeq 600\,$GHz). The extrapolation from 545 GHz to 600 GHz was made using, for each {\it Planck} blazar, its spectral index between 545 and 857 GHz; since the spectral indices are quite flat, the corrections are small.

The upper panel of Fig.~\ref{fig:col_plot} shows that the 9 known blazars in our H-ATLAS fields and the 9 ERCSC blazars lie in a region in the $\log(S_{500\mu\rm m}/S_{1.4\rm GHz})$ vs $\log(S_{500\mu\rm m}/S_{350\mu\rm m})$ plane fairly well separated from the region populated by the majority of the other sources in our parent sample that have a $S_{500\mu\rm m}/S_{1.4\rm GHz} \geq 10$, typical of star forming galaxies. To illustrate this further, we have included in this panel a thick dashed line that represents the colour track as a function of redshift of the prototype starburst galaxy Arp~220. Objects with $S_{500\mu\rm m}/S_{1.4\rm GHz} < 10$ likely have radio emission dominated by an AGN component, that since we are using low-frequency (1.4 GHz) radio observations, is expected to be mostly steep-spectrum (i.e. not associated to blazars).

The $S_{500\mu\rm m}/S_{1.4\rm GHz}$ ratios of blazars are $\le 2.4$ and their $S_{500\mu\rm m}/S_{350\mu\rm m}$ colours are redder than average. This upper limit to $S_{500\mu\rm m}/S_{1.4\rm GHz}$ ratio for blazars comes from the ERCSC objects, while the H-ATLAS ones, which may be thought to be more relevant here, have significantly lower ratios. But we must beware of a selection effect. Since the BZCAT blazars are mostly radio selected (with important additions from X-ray and $\gamma$-ray selections), relatively radio-faint objects, such as possible H-ATLAS blazars with high $S_{500\mu\rm m}/S_{1.4\rm GHz}$ ratios, are easily missed by previous blazar searches. In fact, as will be seen in the following, our candidate H-ATLAS blazars have very limited photometric data. The situation is different for the much (generally more than 10 times) brighter ERCSC blazars for which we likely have the full range of $S_{500\mu\rm m}/S_{1.4\rm GHz}$ ratios. We have therefore adopted as the limiting ratio, the maximum value for ERCSC blazars. Indeed the main strength of the sub-mm selection compared to the traditional ones is its unique efficiency in selecting objects with high $S_{500\mu\rm m}/S_{1.4\rm GHz}$ ratios. This means that even with the adopted limiting ratio we may miss some blazars. For this reason we plan follow-up observations also of objects with ratios above the chosen limit although we expect that such objects are unlikely to be blazars.

The criterion $S_{500\mu\rm m}/S_{1.4\rm GHz}\le 2.4$  yields 12 new candidate blazars in our parent sample (see Table~\ref{tab:tabla_candidates}). From them, 7 have an angular distance to a FIRST source $>5$ arcsec and a $S_{1.4\rm GHz} > 23\,\rm{mJy}$. Since the surface density of the FIRST sources at this flux density limit is $\sim 8\,\hbox{deg}^{-2}$, the expected number of chance associations for angular distances between 5 and 10 arcsec is $\sim1.7$ and some of these associations may indeed be spurious. Thus almost of all the objects in our parent sample are likely to be pure star-forming galaxies or radio galaxies in star-forming hosts as studied in \cite{Jarvis2010} and \cite{Hardcastle2010} respectively.

As a second diagnostic tool, we have built the  $(r-z)$ vs. $(u-r)$  colour-colour diagram (lower panel of Fig.~\ref{fig:col_plot}) for known blazars by cross-correlating the SDSS DR8\footnote{http://www.sdss3.org/dr8/} catalogue \citep{DR8} with the 
BZCAT, using a search radius of $0.5$ arcsec. The surface density of DR8 sources is $\simeq 3.22\times 10^4 \,\hbox{deg}^{-2}$ so that the probability of a chance association within $0.5$ arcsec is $\simeq 2\times 10^{-3}$. Since there are $\simeq 1800$ BZCAT sources within the DR8 area, we expect $\simeq 3.5$ chance associations. We get 1600 matches. They are mostly concentrated along a relatively narrow strip in the $(r-z)$ vs. $(u-r)$ plane. Seven out of the 12 new candidates (\#~11, 15, 16, 17, 19, 20, and 21) 
lie within or very close to this strip, although the \textsl{Herschel}/SPIRE colours of \#~16 and 17 may be indicative of contamination by dust emission in the host galaxy. Objects \#~12, 14, and 18 
are at, or somewhat above, the red end of the strip, suggesting that their optical colours may be moderately contaminated by the host galaxy. Hints of host galaxy contamination come also from the \textsl{Herschel}/SPIRE colours of two of these objects (\#~12 and 14). 
As illustrated by the colour tracks in the lower panel of Fig.~\ref{fig:col_plot}, very red colours are indicative of contamination by a passive early-type host galaxy. Indications that colours redder than $(u-r)=1.4$ likely imply host galaxy contamination were also reported by \cite{Massaro2012}. The sub-mm to optical photometric data on one of the two outliers with blue $(u-r)$ colours (object \#~10) are strongly indicative of a star-forming galaxy. The other blue outlier (object \#~13) is anomalously faint in the $r$-band (compared to neighbour bands); its $(r-z)$ versus $(u-r)$ colours may thus not be a good diagnostic. Some of the new Blazar candiates have optical colours consistent with QSOs.

A third diagnostic tool, based on the [3.4][4.6][12]$\,\mu$m colour--colour diagram using WISE \citep{Wright2010} magnitudes, was proposed by \citet{Massaro2011}. Magnitudes for the 3 WISE channels are available\footnote{http://irsa.ipac.caltech.edu/Missions/wise.html} for 7  H-ATLAS catalogued blazars (the exceptions are \#~7 and 9) and for 10 H-ATLAS candidates (the exceptions are \#~20 and 24). However objects \#~9 and 20 had magnitudes listed in the Preliminary Release of WISE data. The positions of H-ATLAS confirmed or candidate blazars in this diagram are identified, in Fig.~\ref{fig:WISE_diag}, by their ID numbers (see Tables~\ref{tab:tabla_confirm8} and \ref{tab:tabla_candidates}). The catalogued H-ATLAS blazars for which WISE photometry is available plus the candidates \#~10, 16, and 20 lie in the upper part of `blazar strip', i.e. in the region occupied by flat-spectrum radio quasars (although \#~16 is at the border). Four of our blazar candidates (\#~12, 13, 18, and 21)  have WISE colours typical of starburst galaxies, and two (\#~17 and 19) are in the region populated by LINERs/obscured AGNs\footnote{http://wise2.ipac.caltech.edu/docs/release/prelim/expsup/fig\-ures/sec2\_2f16\_annot.gif}.

\subsection{Follow-up observations in radio}\label{sec:Medicina}

We have performed follow-up observations with the Me\-di\-ci\-na 32m single-dish telescope at 5 GHz of 6 blazar candidates.  Observations were carried out between August 1st -16th 2011  in the OTF scan mode  \citep{magnum07,righini08} applying the observing strategy, the data reduction procedures and the software tools developed for the SiMPlE project \citep{procopio11}. Twenty-five hours were allocated for 5 GHz observations with a single feed receiver and a 150 MHz-wide band. The 37 arcmin length of each scan ($\sim 5\times\hbox{HPBW}$) was  covered at 3 arcmin/s in 12.5 s. A sampling rate of 40 ms allowed us to obtain 60 samples/beam. The scans were centred on the FIRST positions.

Given the FIRST flux densities we do not expect to see any of our targets in a single scan, which, in optimal observing conditions, should hence appear as a linear baseline that corresponds to the off-source zero level of the signal. Along the scans, Gaussian noise shows up as amplitude fluctuations. However, cloudy weather, the presence of random, but unfortunately not rare, contributions by radio frequency interference (RFI) or digital noise heavily affected portions of the data. These disturbances give rise to bumpy baselines or spike-like features. All the scans that do not show a linear profile or show such features were removed before running the data reduction pipeline.

We have used the OTF Scan Calibration-Reduction pipeline \citep[OSCaR,][]{procopio11,righini11} to get the conversion factor from raw data counts measured on calibrators to their known flux densities, to determine the component of the flux density error due to calibration, and to calculate the source flux densities and their errors. 3C295 and 3C286 were used as flux density calibrators. One out of the six targeted sources was not detected because of technical problems. The 5 GHz flux densities for the other 5 sources were measured with signal to noise ratios greater than 6. The results are given in Table~\ref{tab:med}. Source \#\,14 turned out to have a steep 1.4--5 GHz spectral index and is optically identified as a galaxy. For the other 4 sources (\# 10, 11, 13, and 21) our observations indicate a flat spectral index, consistent with them being blazars. We caution, however, that the low frequency spectral index is not necessarily a conclusive indication for or against a blazar classification since the low frequency radio emission frequently comes from a different component. Additional flux density measurements up to mm wavelengths are necessary to assess the nature of candidate blazars. Such observations are being planned.

\begin{table}
\begin{center}
\begin{tabular}{cccccc}
\hline ID & RA ($^{\circ}$) & Dec. ($^{\circ}$) & $S_{1.4\rm GHz}$  &  $S_{5\rm GHz}$  & $\sigma_{S}$ \\
\hline
10 & 218.522& -0.385& 59.92 & 43.03  & 5.25 \\
11 & 220.332& 0.420 & 57.32 & 38.72  & 3.19 \\
13 & 222.376& 2.610 & 50.09 & 46.93 & 4.30 \\
14 & 174.092& 0.815 & 151.74 & 37.21  & 3.93 \\
21 & 222.406& 0.693 & 51.93 & 55.20  & 9.09 \\
\hline
\end{tabular}
\caption{Flux densities (mJy) at 5 GHz and their errors ($\sigma_{S}$) measured with the Medicina radio telescope for 5 blazar candidates. The FIRST 1.4 GHz flux densities (mJy) are also shown, for comparison.}\label{tab:med}
\end{center}
\end{table}

\begin{figure*}
\begin{center}
\includegraphics[width=0.32 \textwidth]{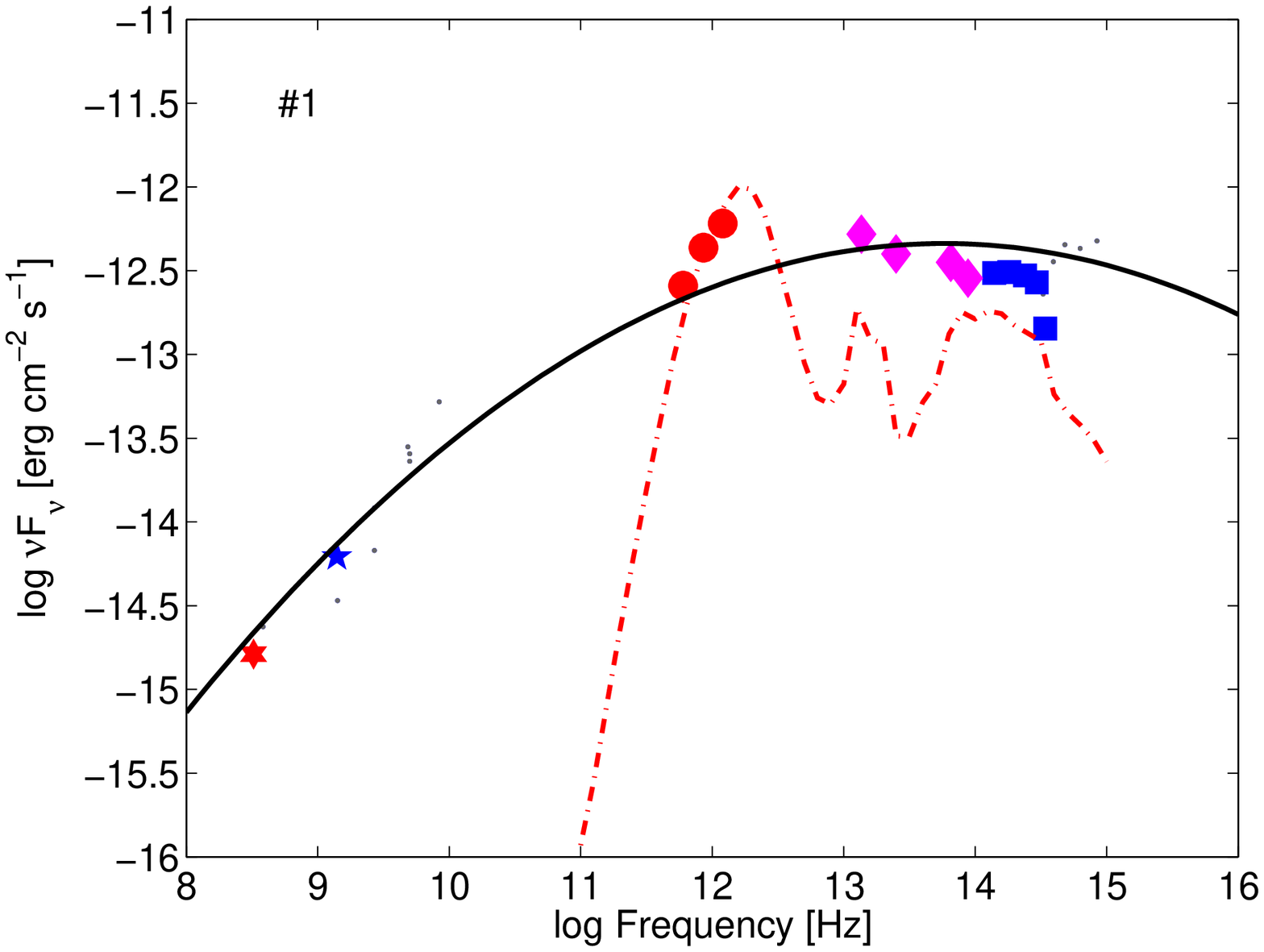}
\includegraphics[width=0.32 \textwidth]{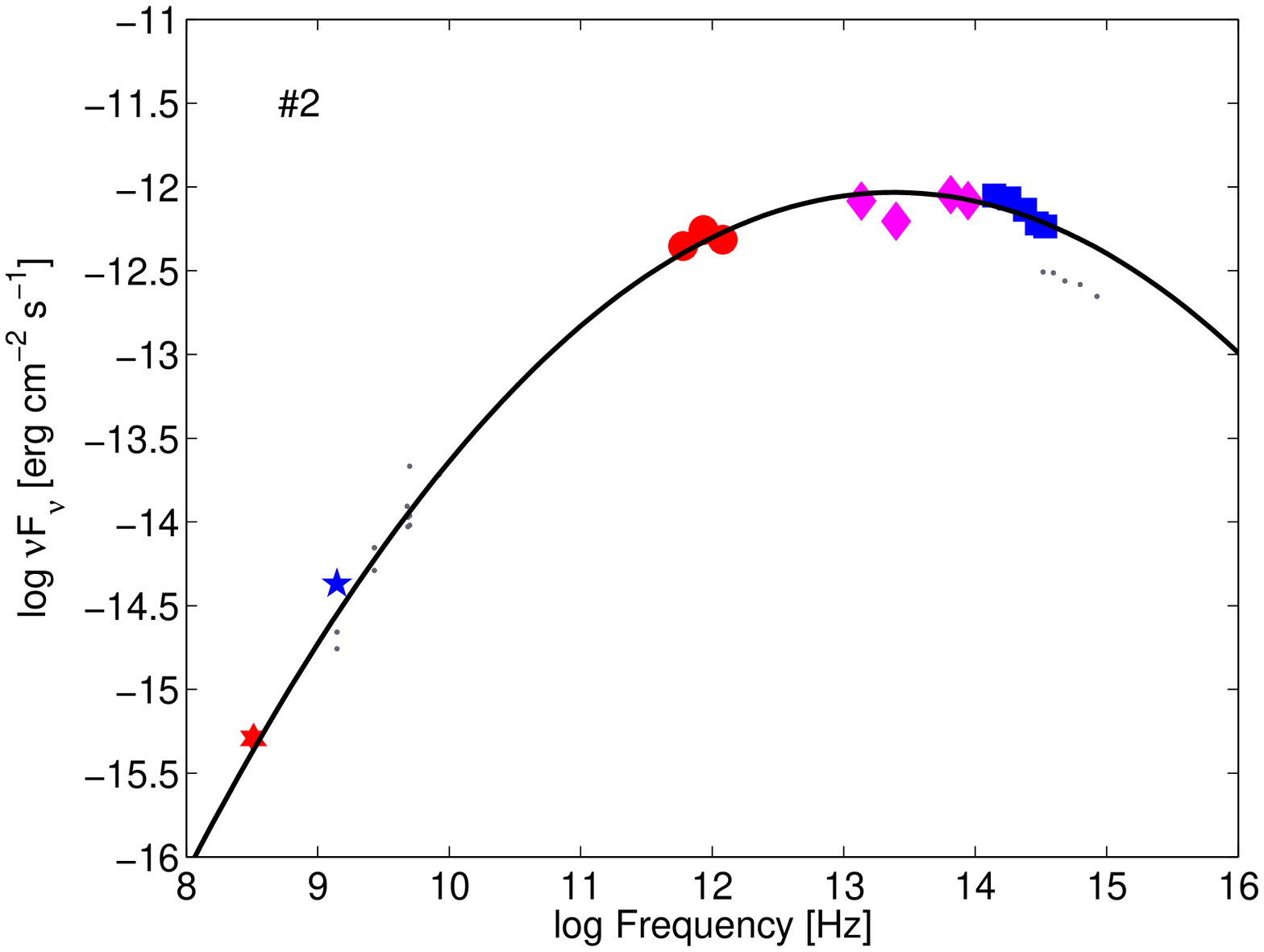}
\includegraphics[width=0.32 \textwidth]{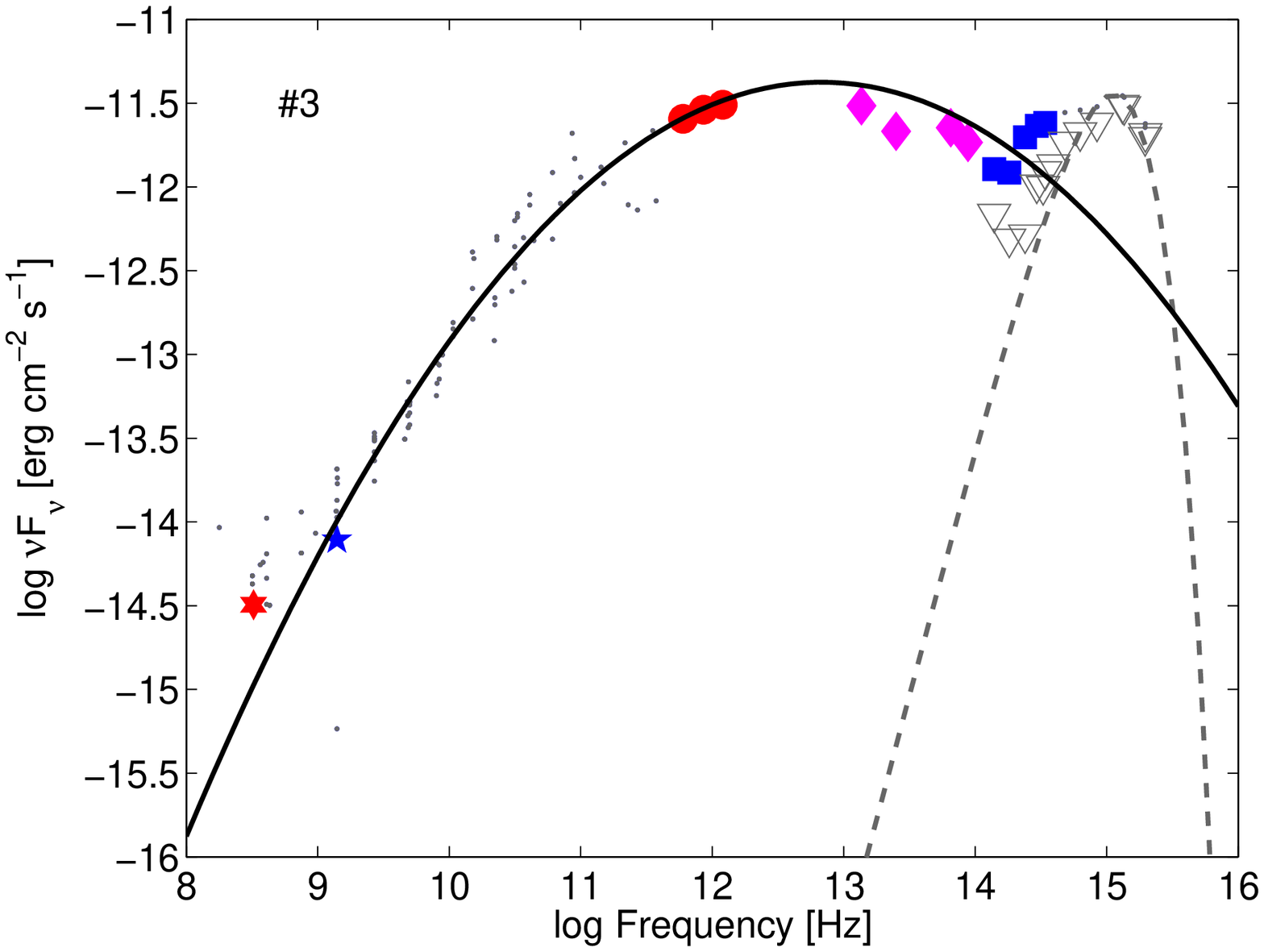}
\includegraphics[width=0.32 \textwidth]{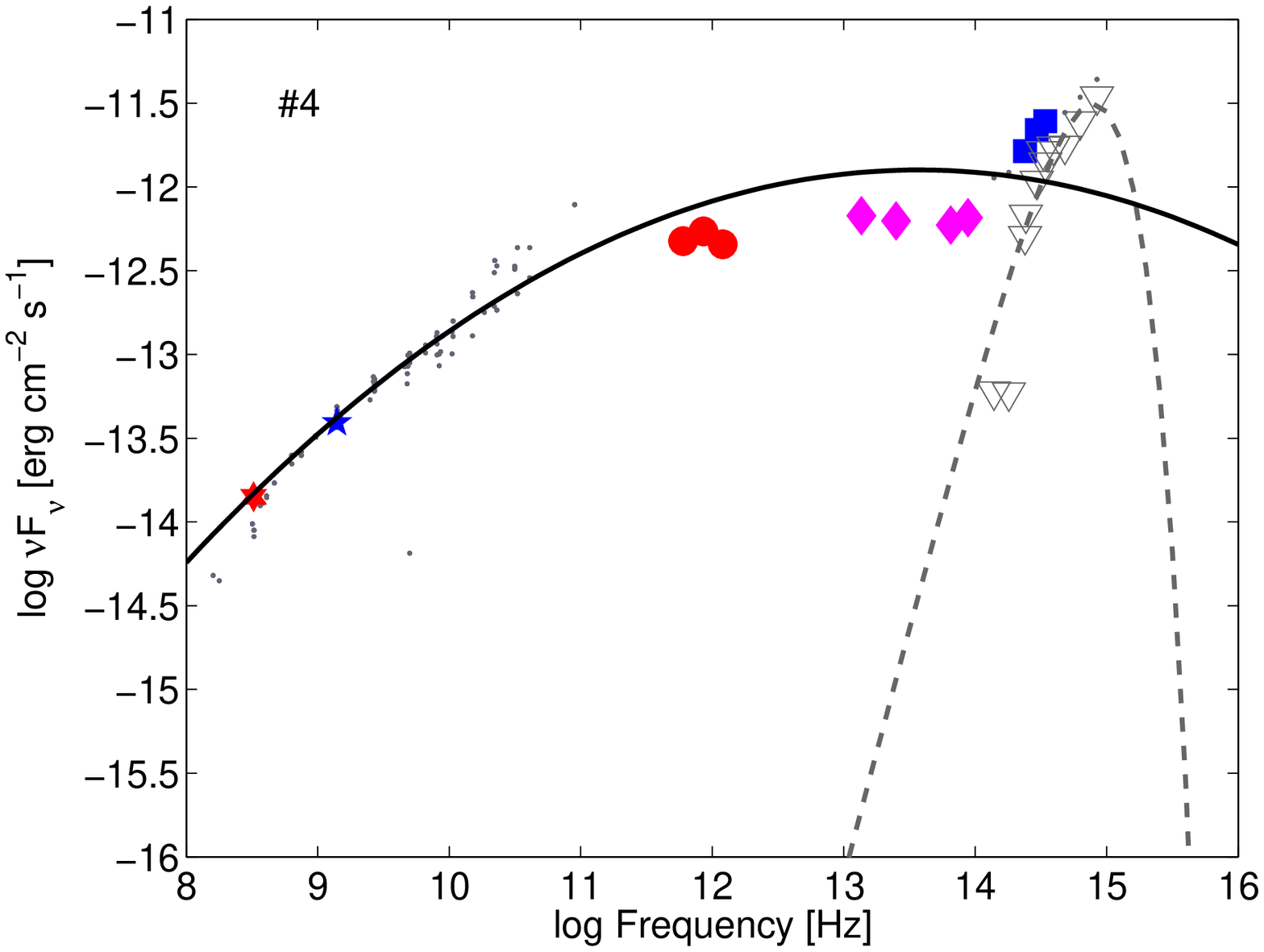}
\includegraphics[width=0.32 \textwidth]{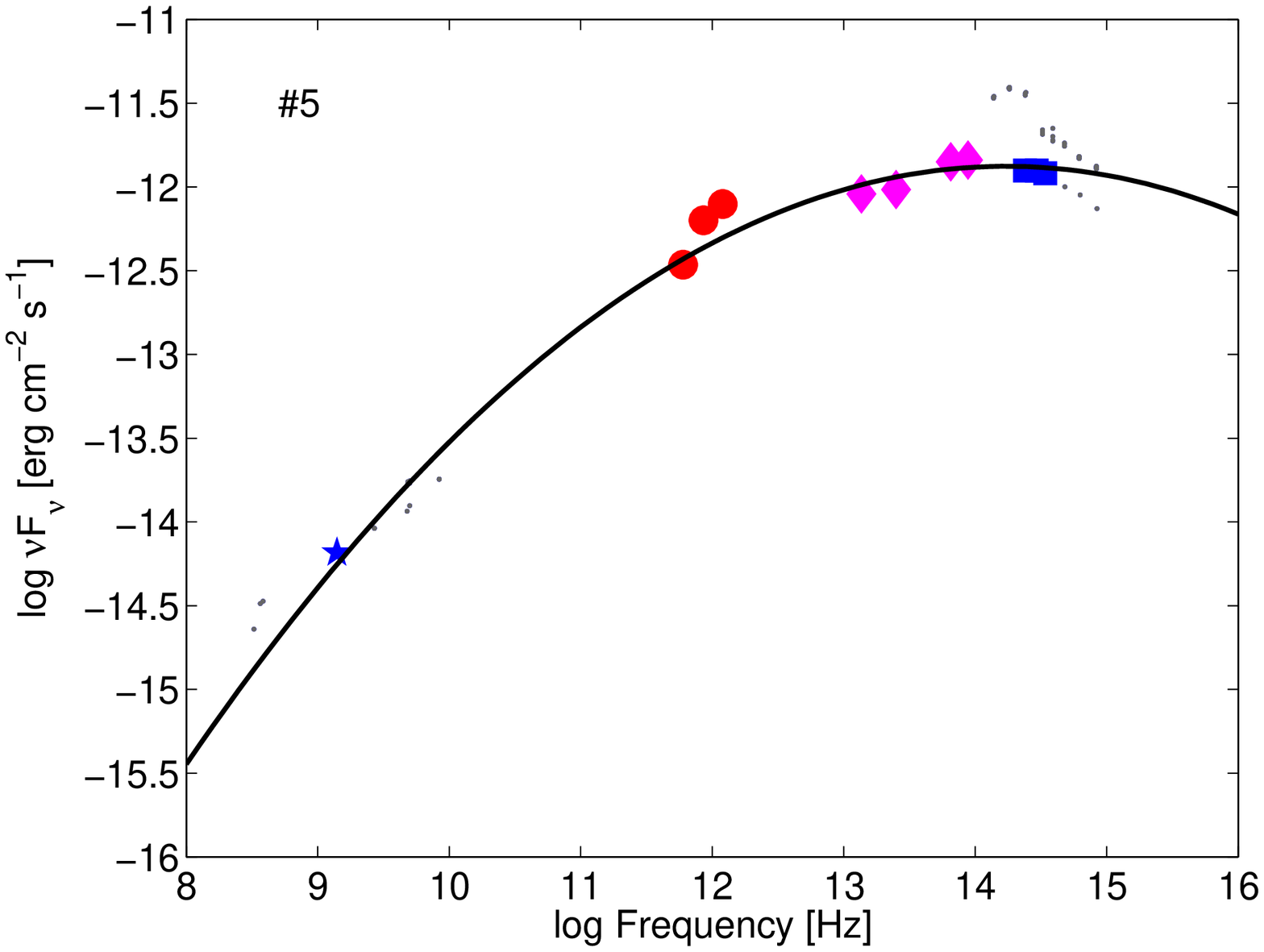}
\includegraphics[width=0.32 \textwidth]{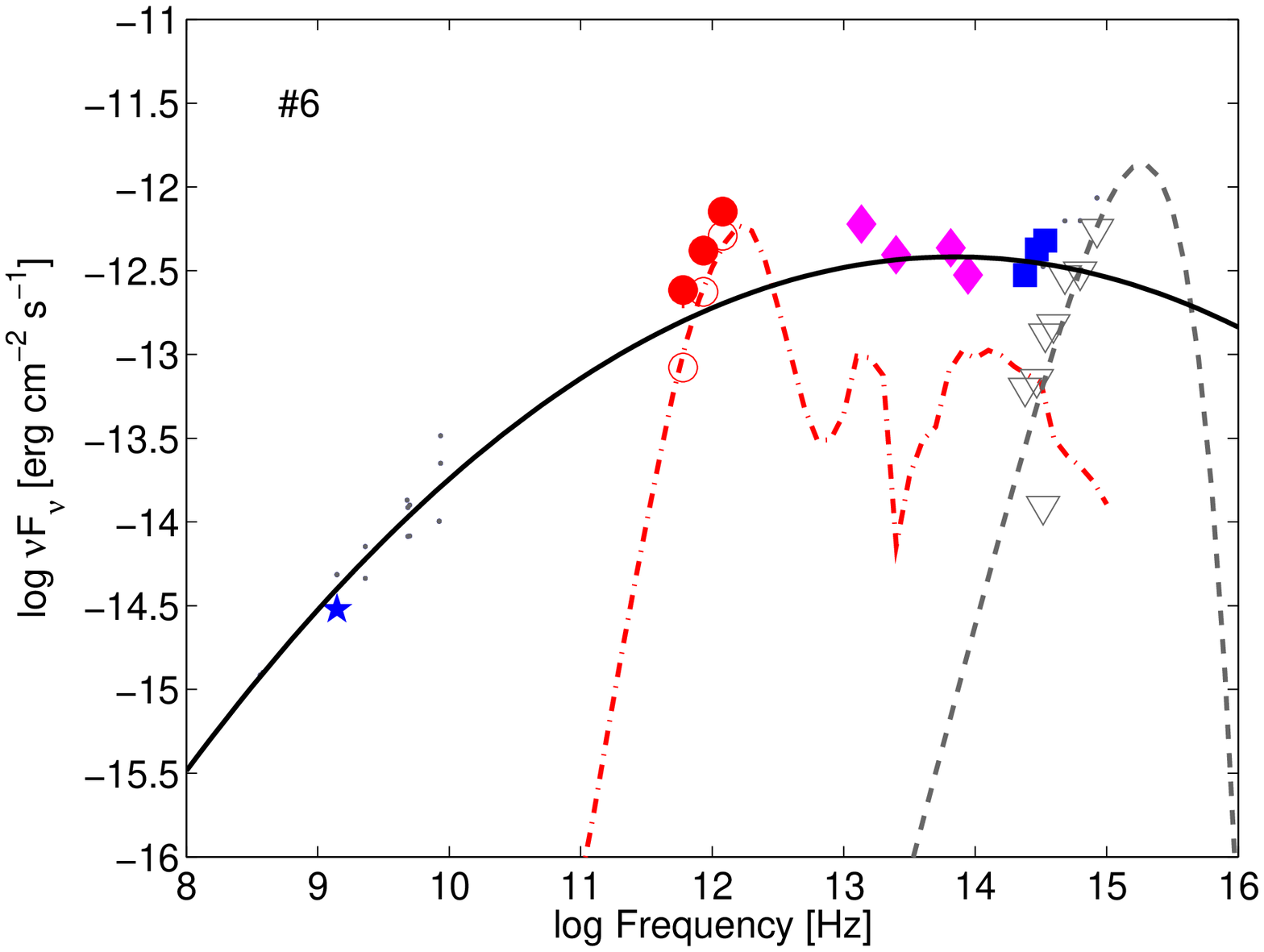}
\includegraphics[width=0.32 \textwidth]{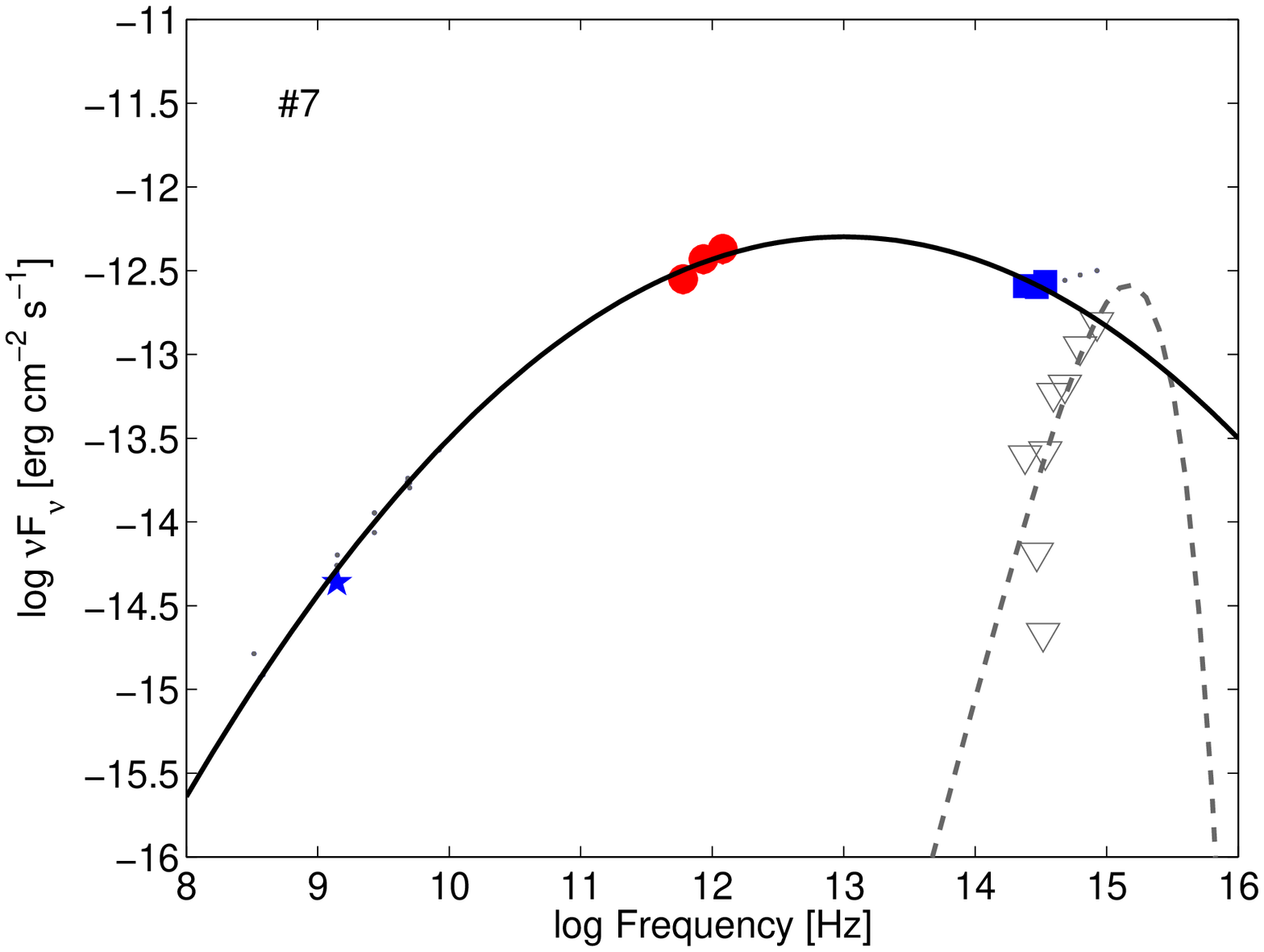}
\includegraphics[width=0.32 \textwidth]{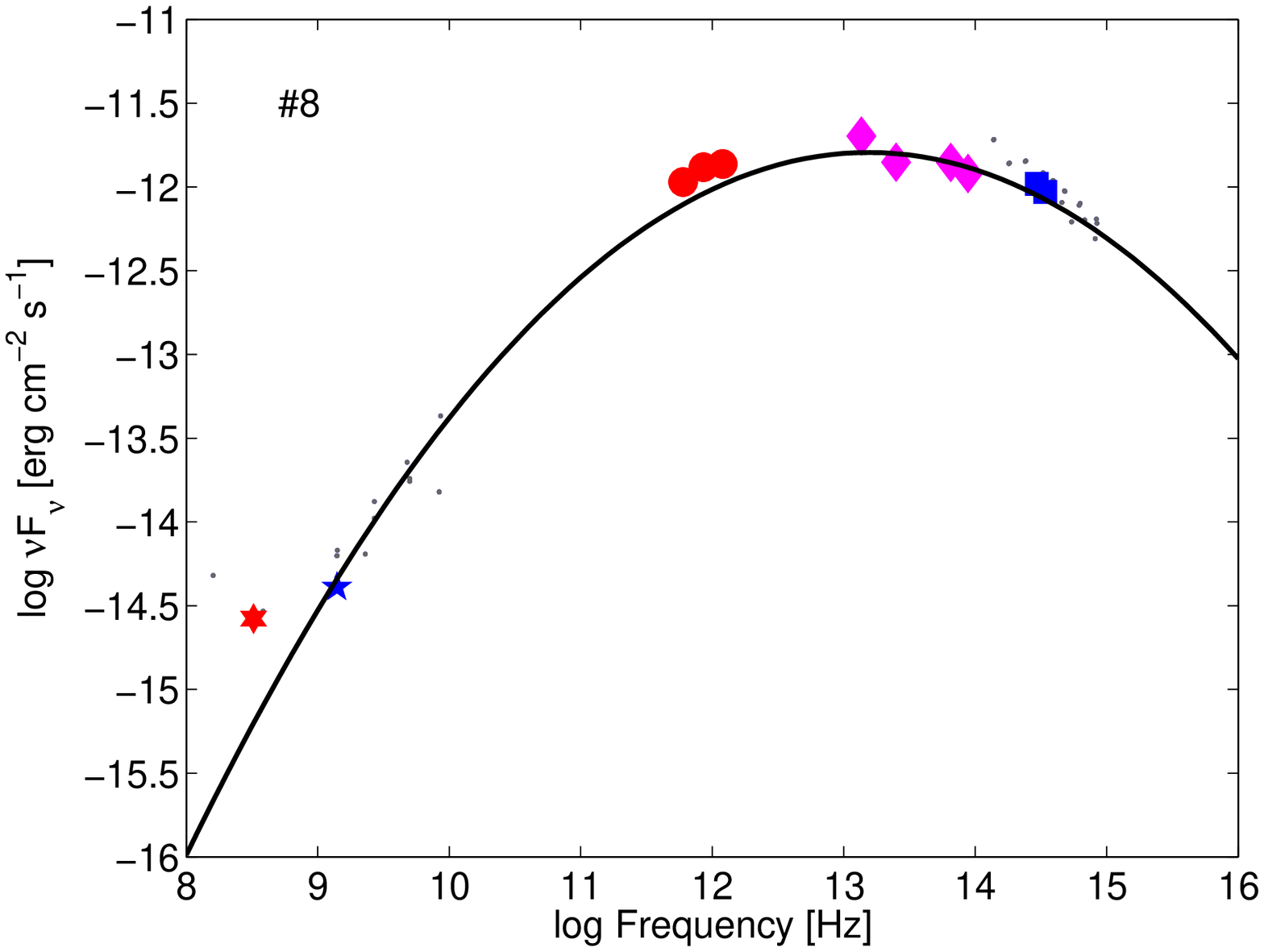}
\includegraphics[width=0.32 \textwidth]{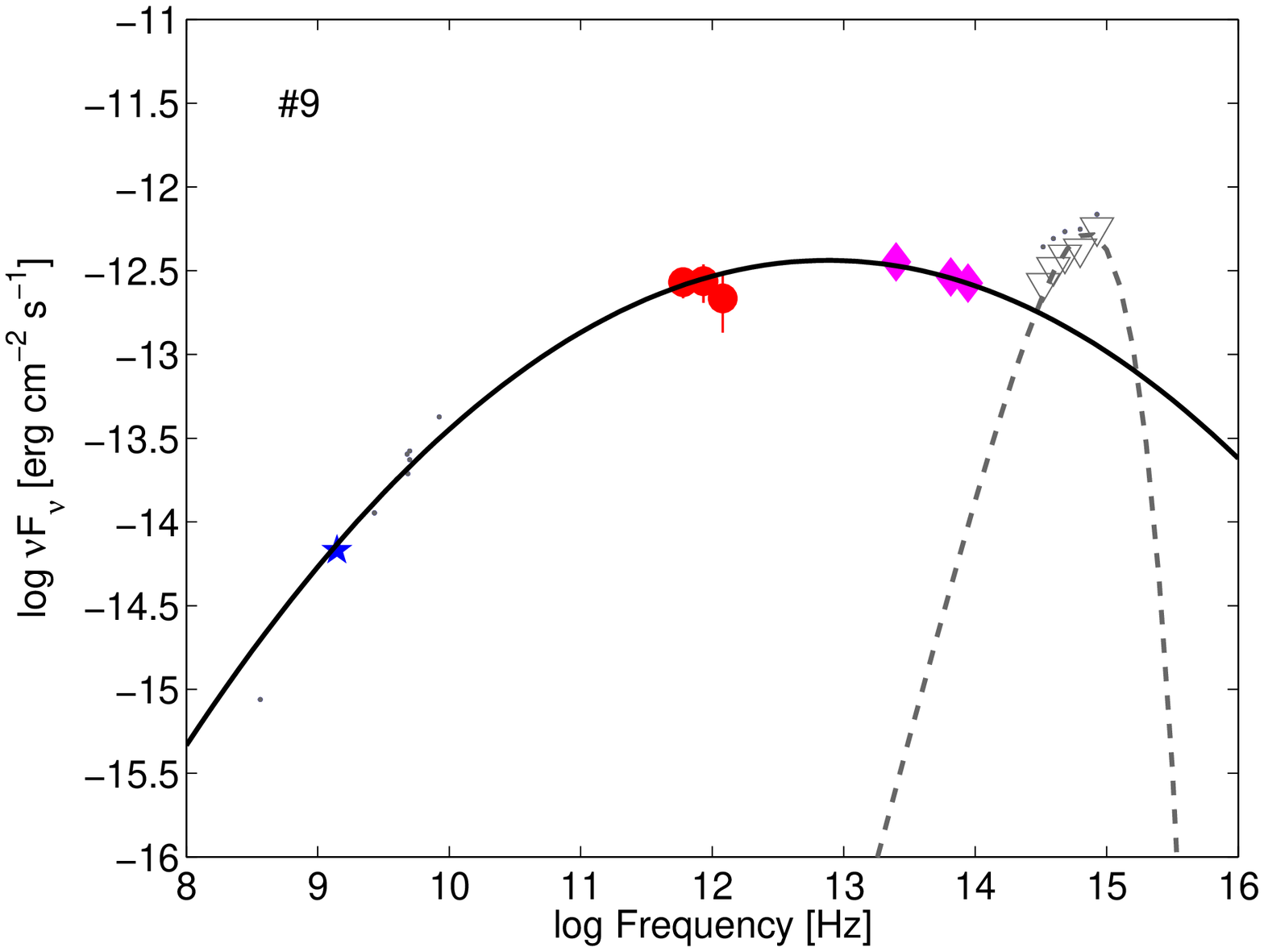}
\end{center}
\caption{SEDs of the 9 catalogued blazars in the H-ATLAS equatorial fields (Table~\protect\ref{tab:tabla_confirm8}). The red circles, magenta diamonds, blue squares, red 6-pointed stars, blue 5-pointed stars and green asterisks are H-ATLAS, WISE, VIKING, GMRT, FIRST and Medicina data, respectively. The other data points are from the NASA/IPAC Extragalactic Database (NED). The downward-pointing triangles are the results of subtracting from the data the fitted second order polynomial (solid lines) representing the synchrotron emission. In panels 1 and 6 the dot-dashed lines show the dusty galaxy template for $\log(L_{\rm dust}/L_\odot)>11.5$ \citep{smith2012}, put at the blazar redshifts ($z=1.123$ and $z=1.173$) respectively. This template is only shown for illustration purposes and no fit to the data has been attempted: since the synchrotron component is highly variable and the data are not simultaneous, fits would have little meaning. The dashed curves on the right side of panels 3, 4, 6, 7, and 9 are black-body spectra representing the possible emission from the accretion disk. Again, these curves are not meant to be fits of the data. } \label{fig:sed_blazars}

\end{figure*}
\begin{figure*}
\begin{center}
\includegraphics[width=0.32 \textwidth]{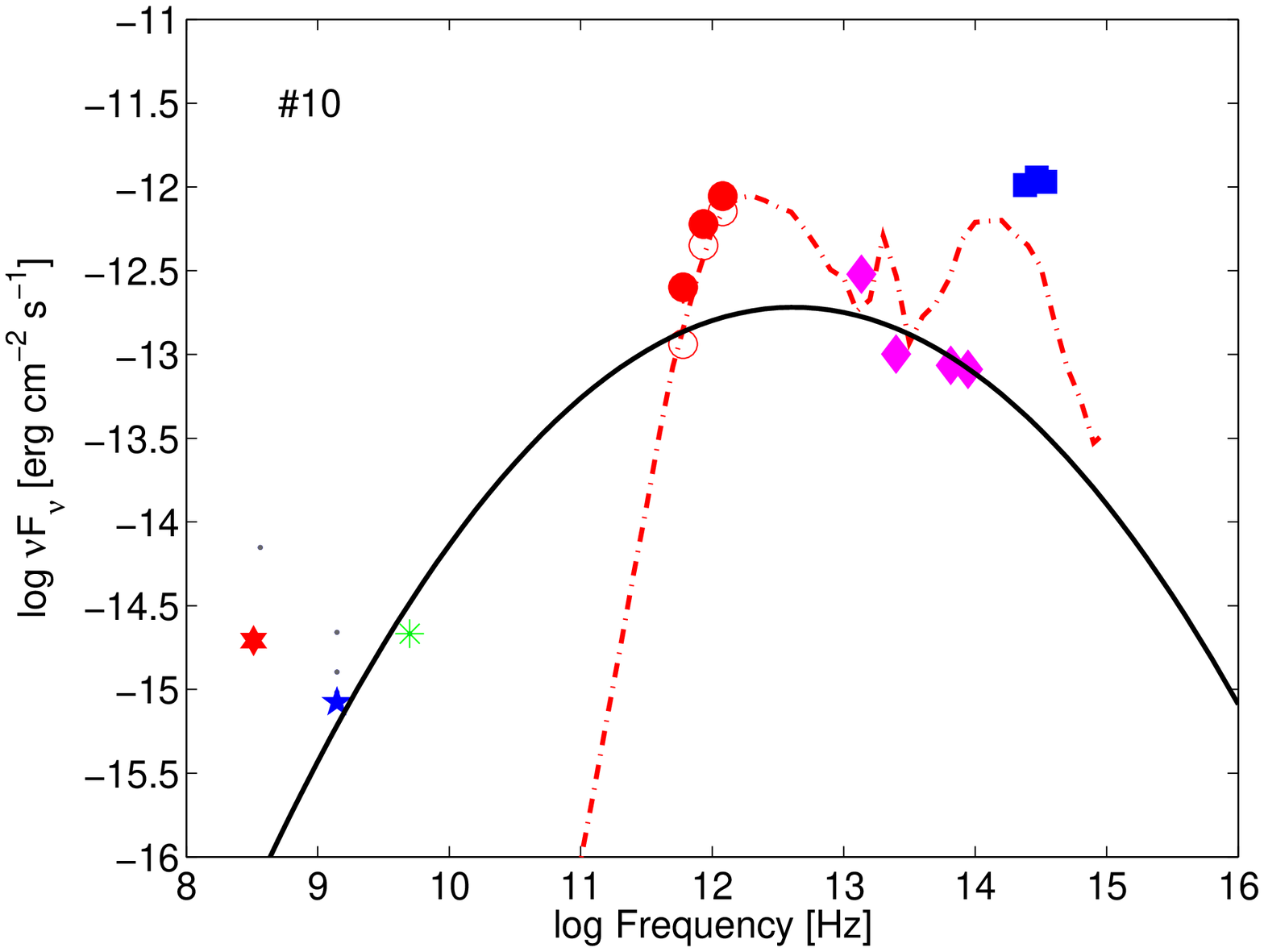}
\includegraphics[width=0.32 \textwidth]{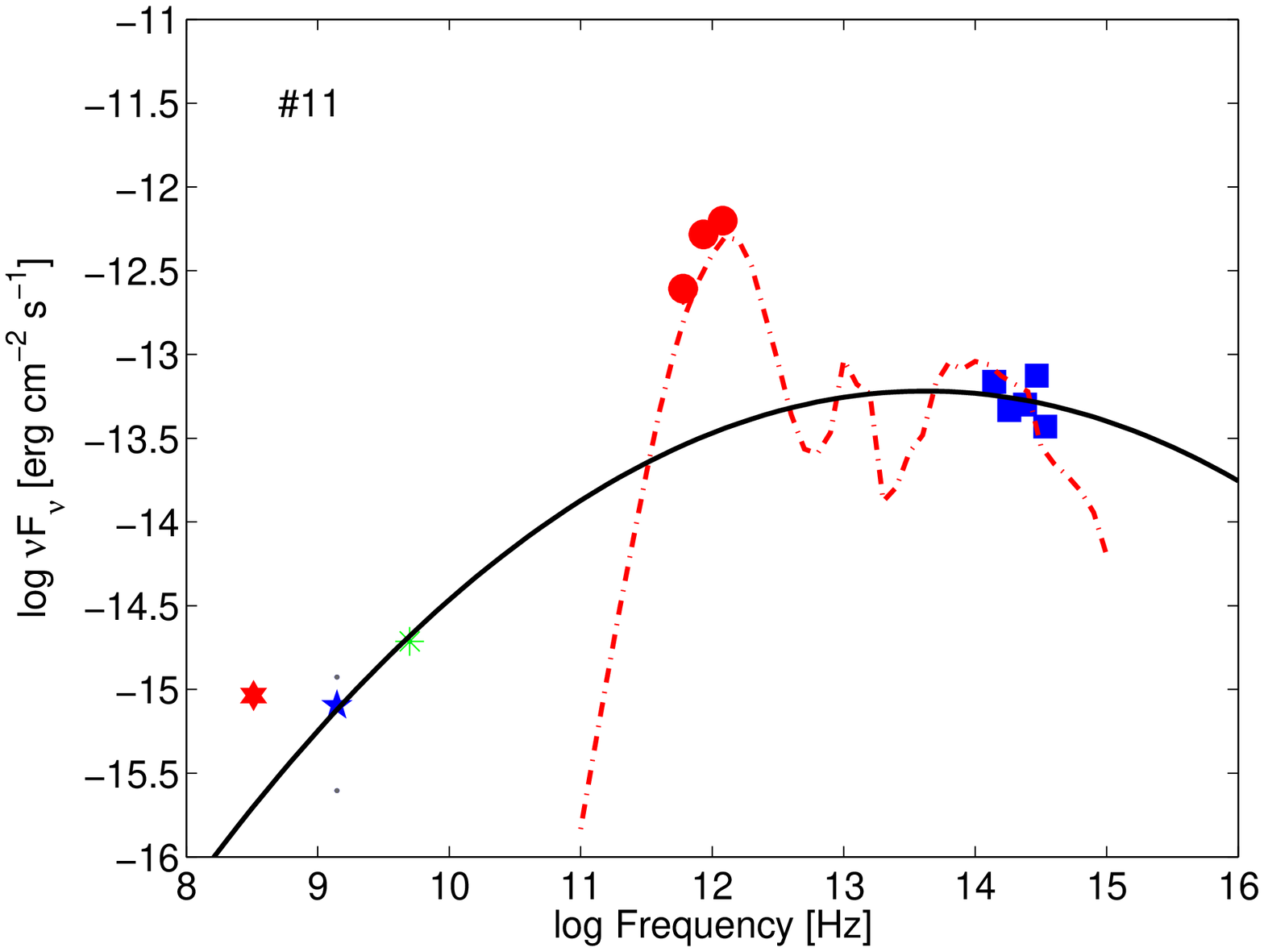}
\includegraphics[width=0.32 \textwidth]{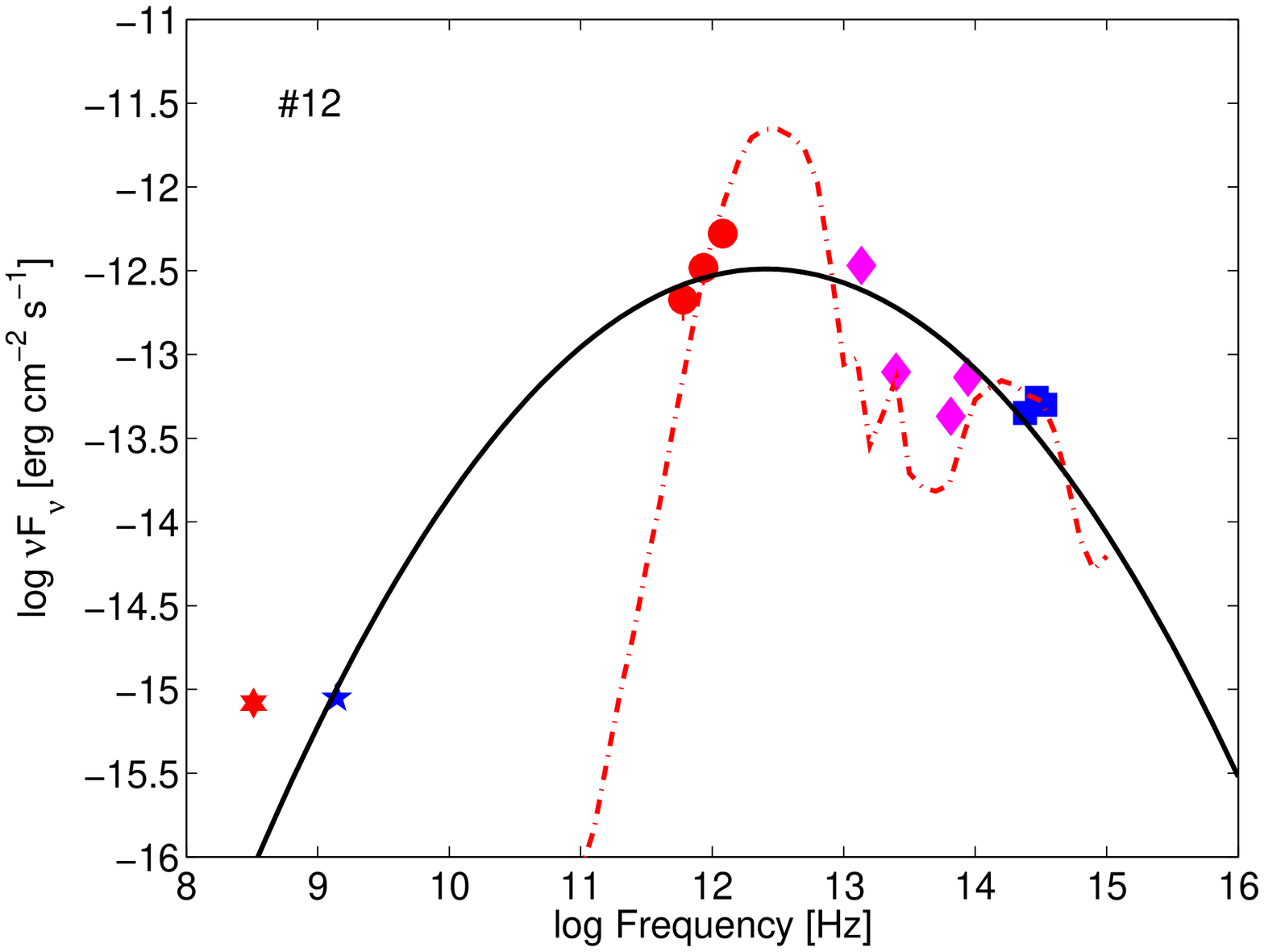}
\includegraphics[width=0.32 \textwidth]{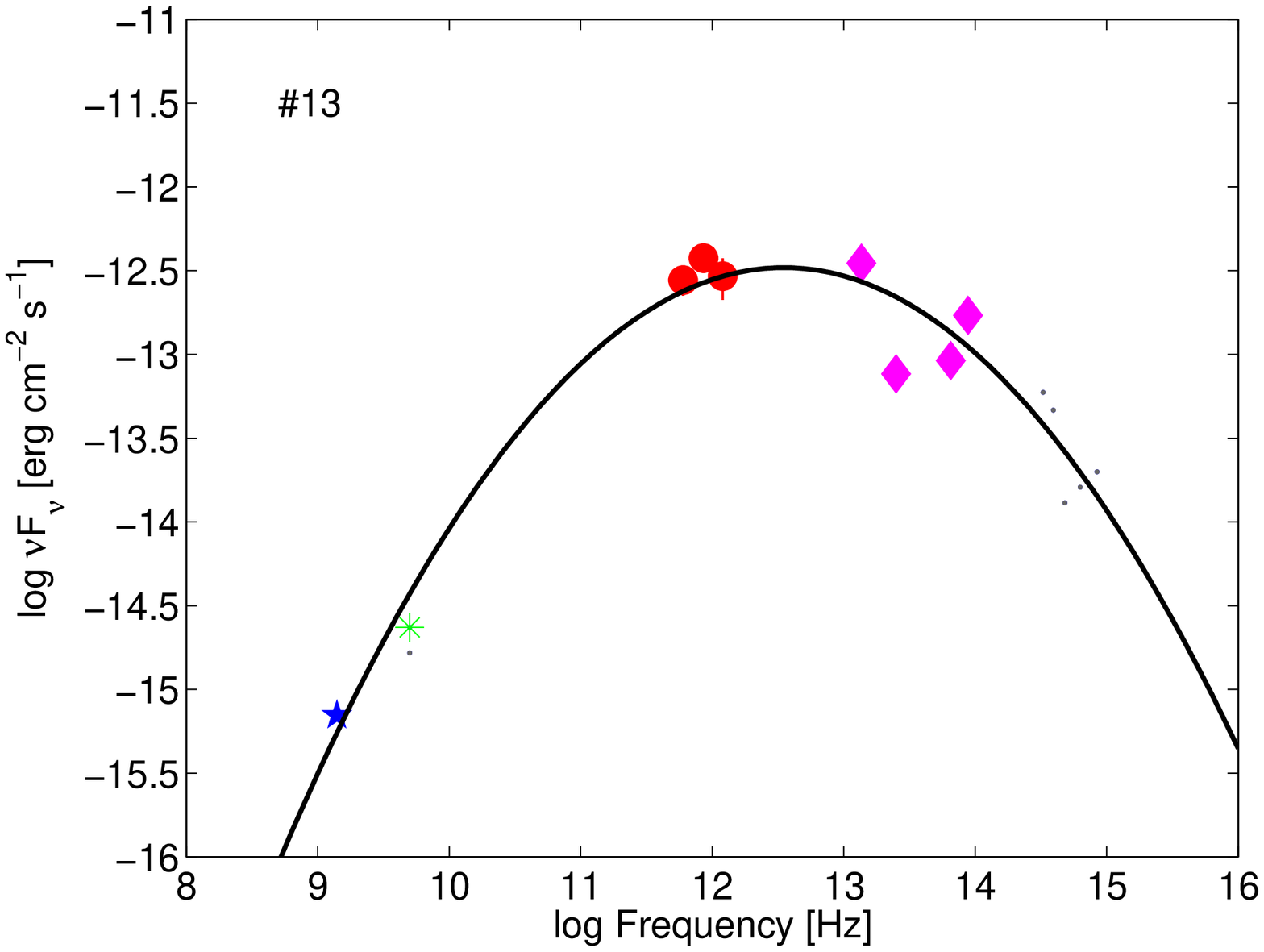}
\includegraphics[width=0.32 \textwidth]{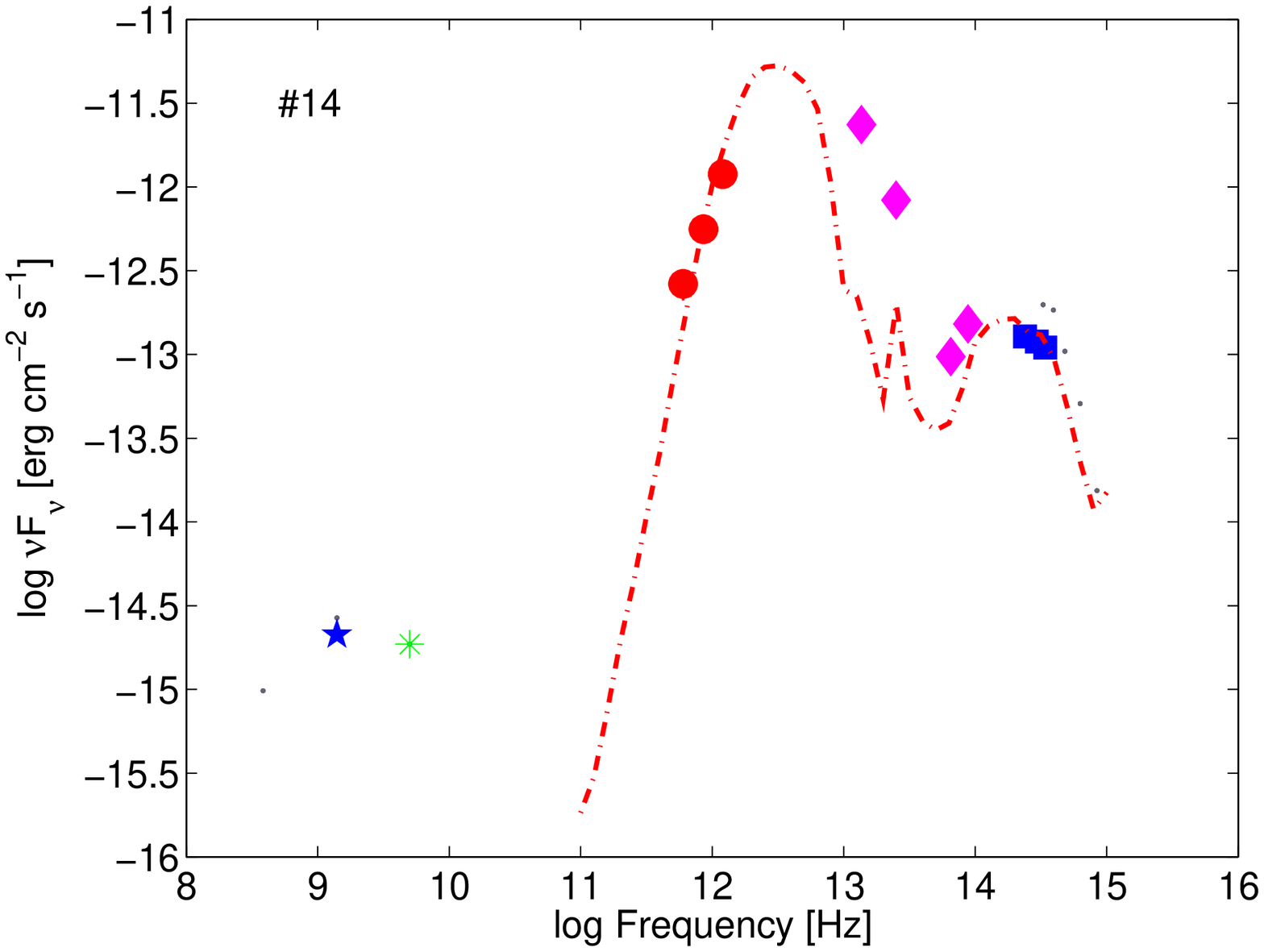}
\includegraphics[width=0.32 \textwidth]{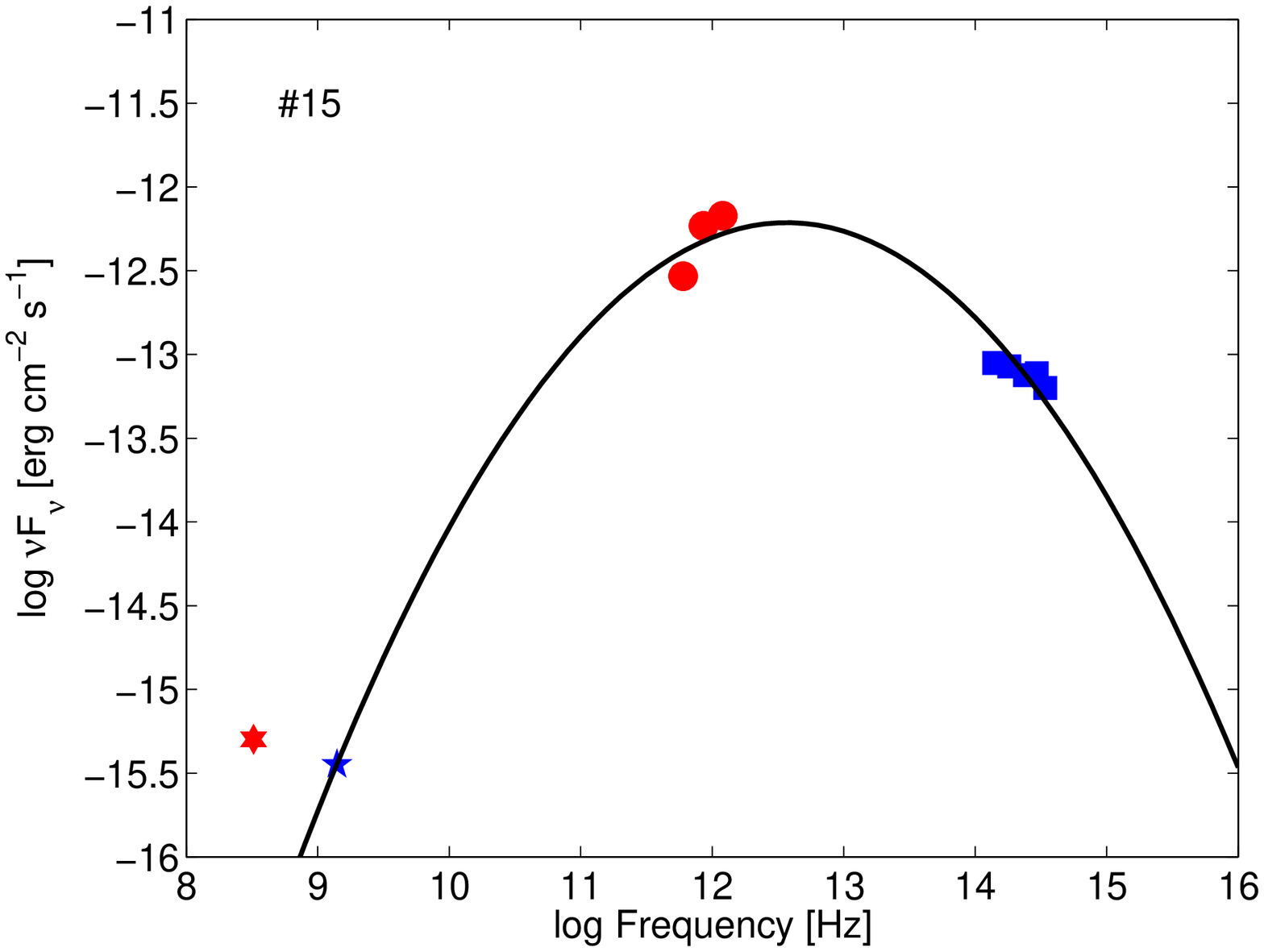}
\includegraphics[width=0.32 \textwidth]{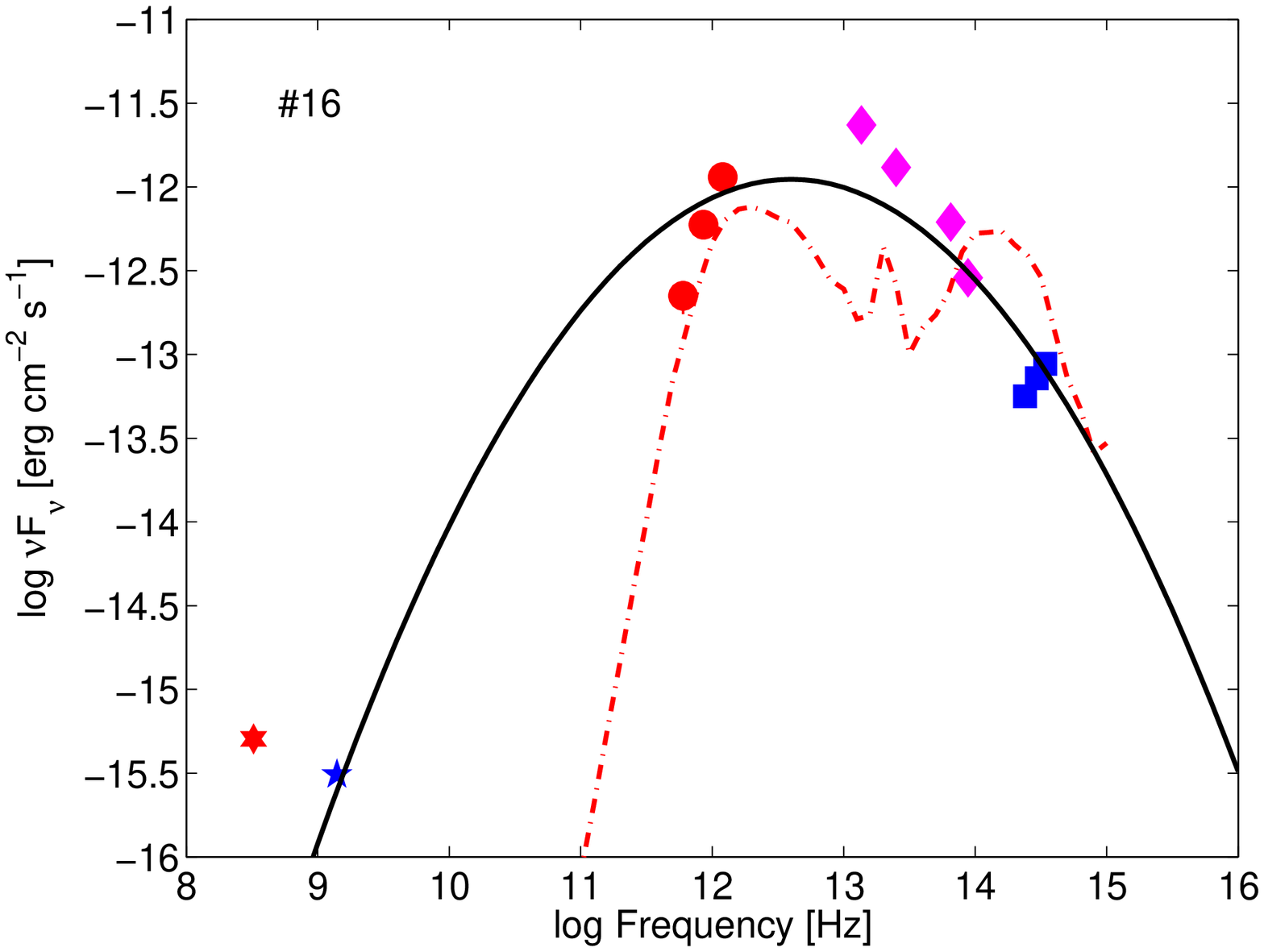}
\includegraphics[width=0.32 \textwidth]{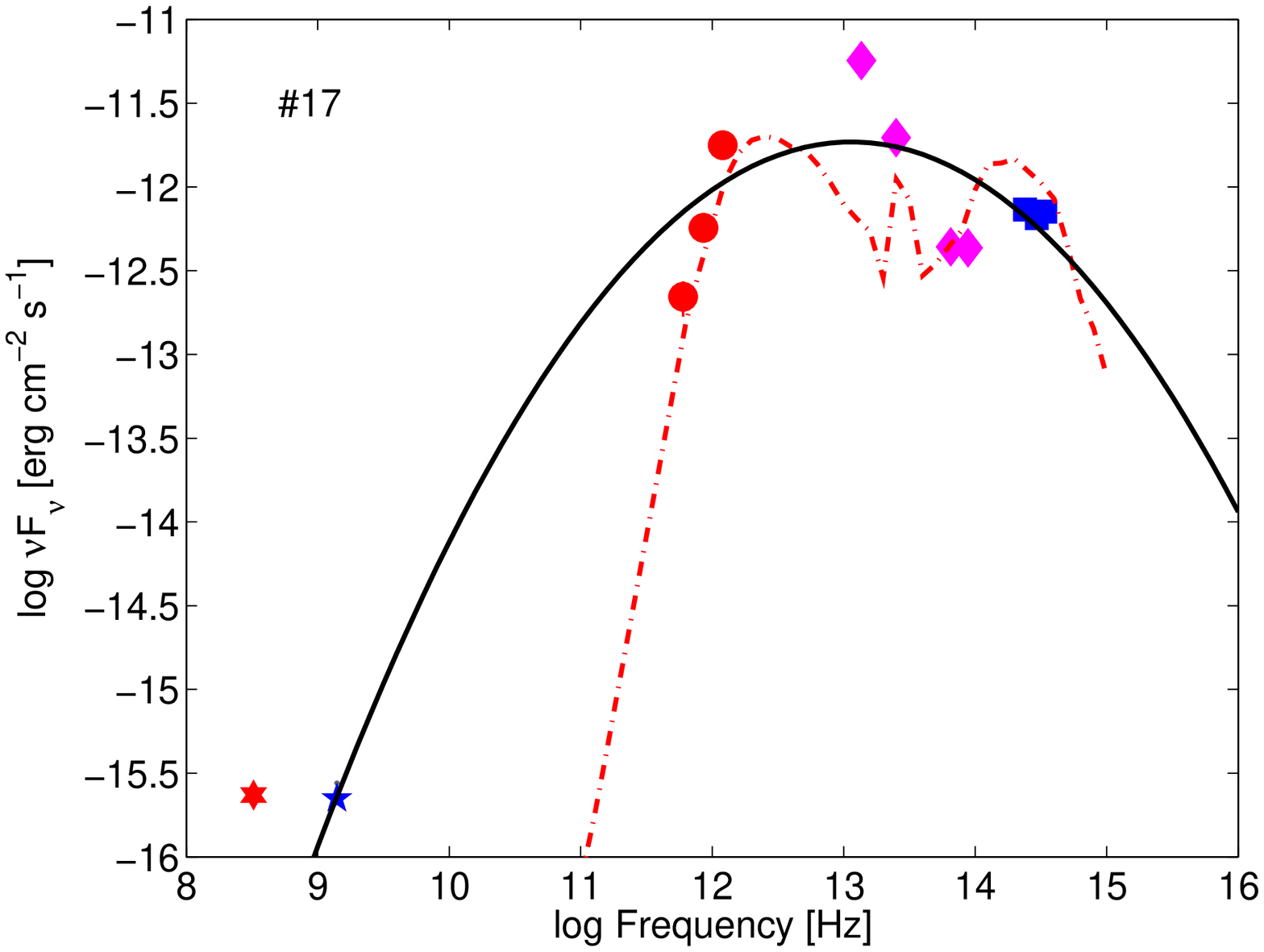}
\includegraphics[width=0.32 \textwidth]{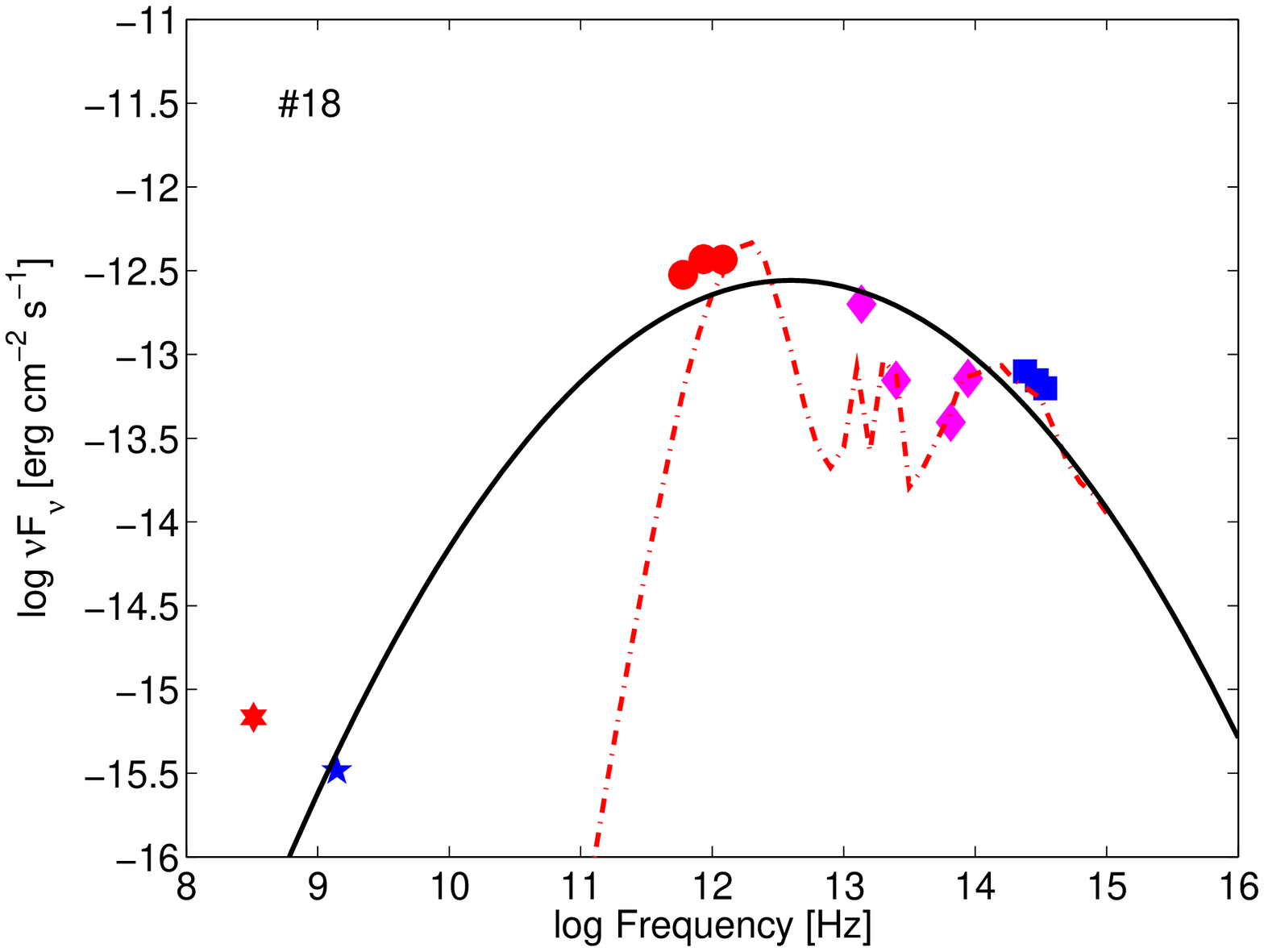}
\includegraphics[width=0.32 \textwidth]{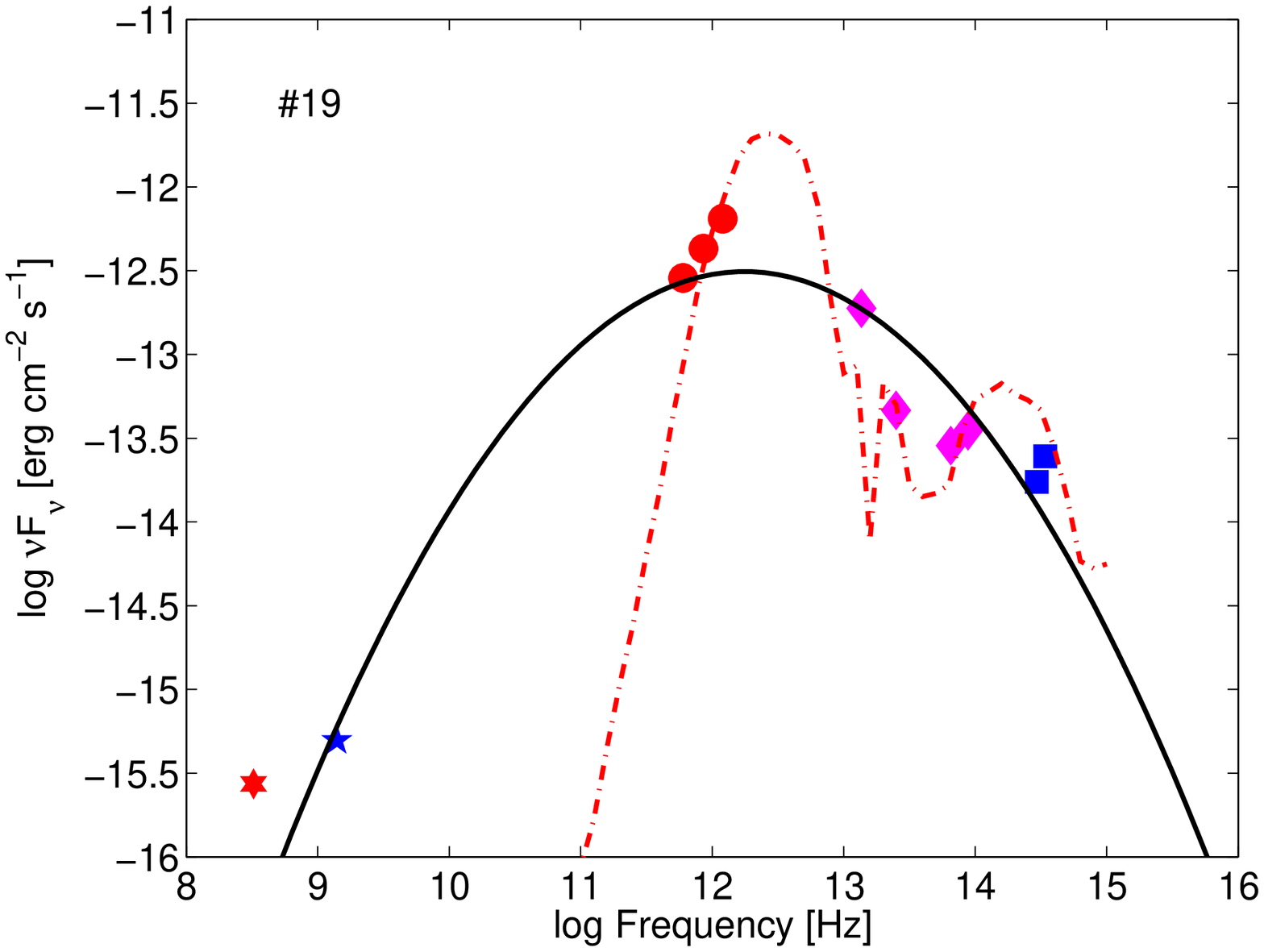}
\includegraphics[width=0.32 \textwidth]{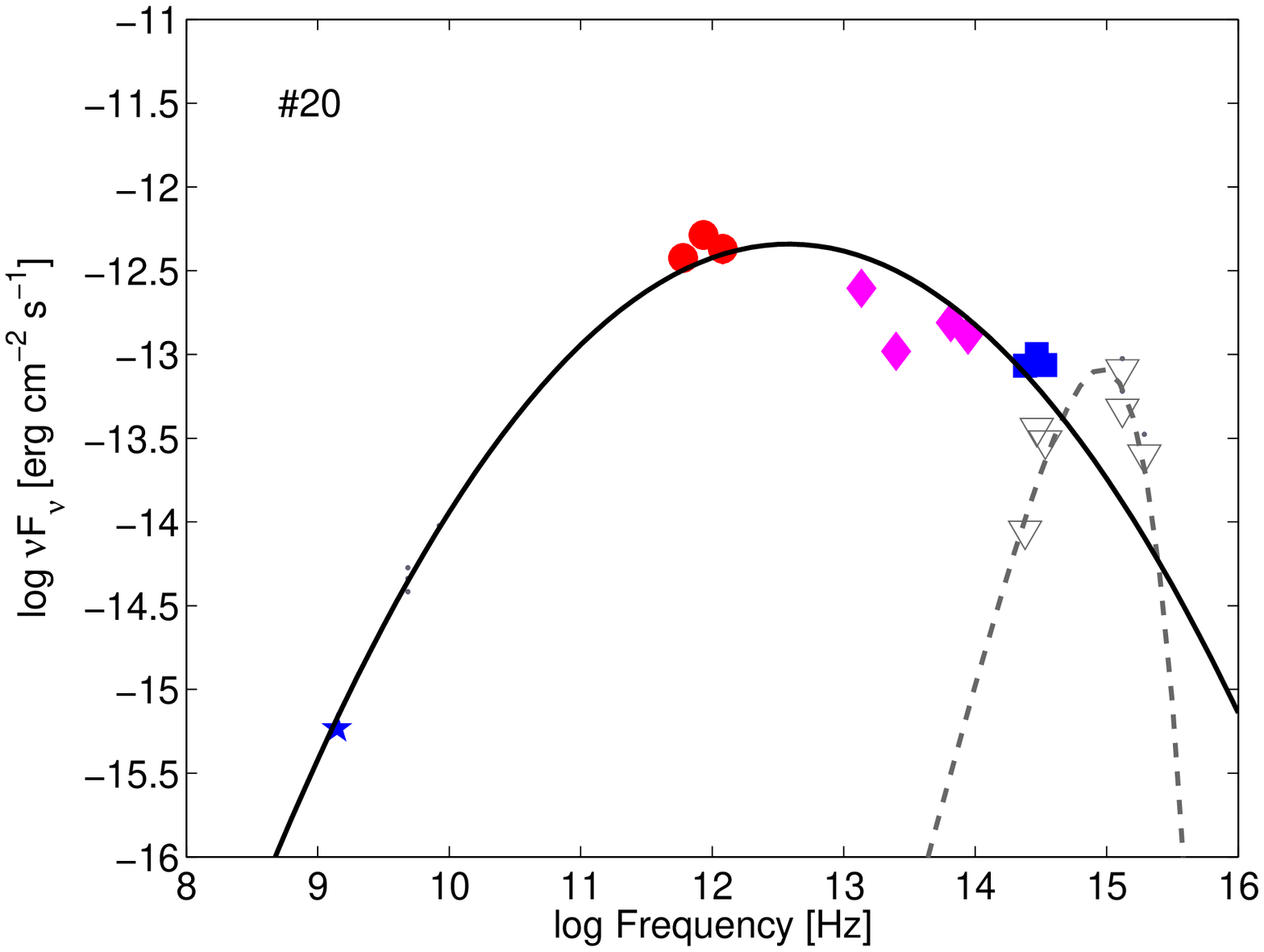}
\includegraphics[width=0.32 \textwidth]{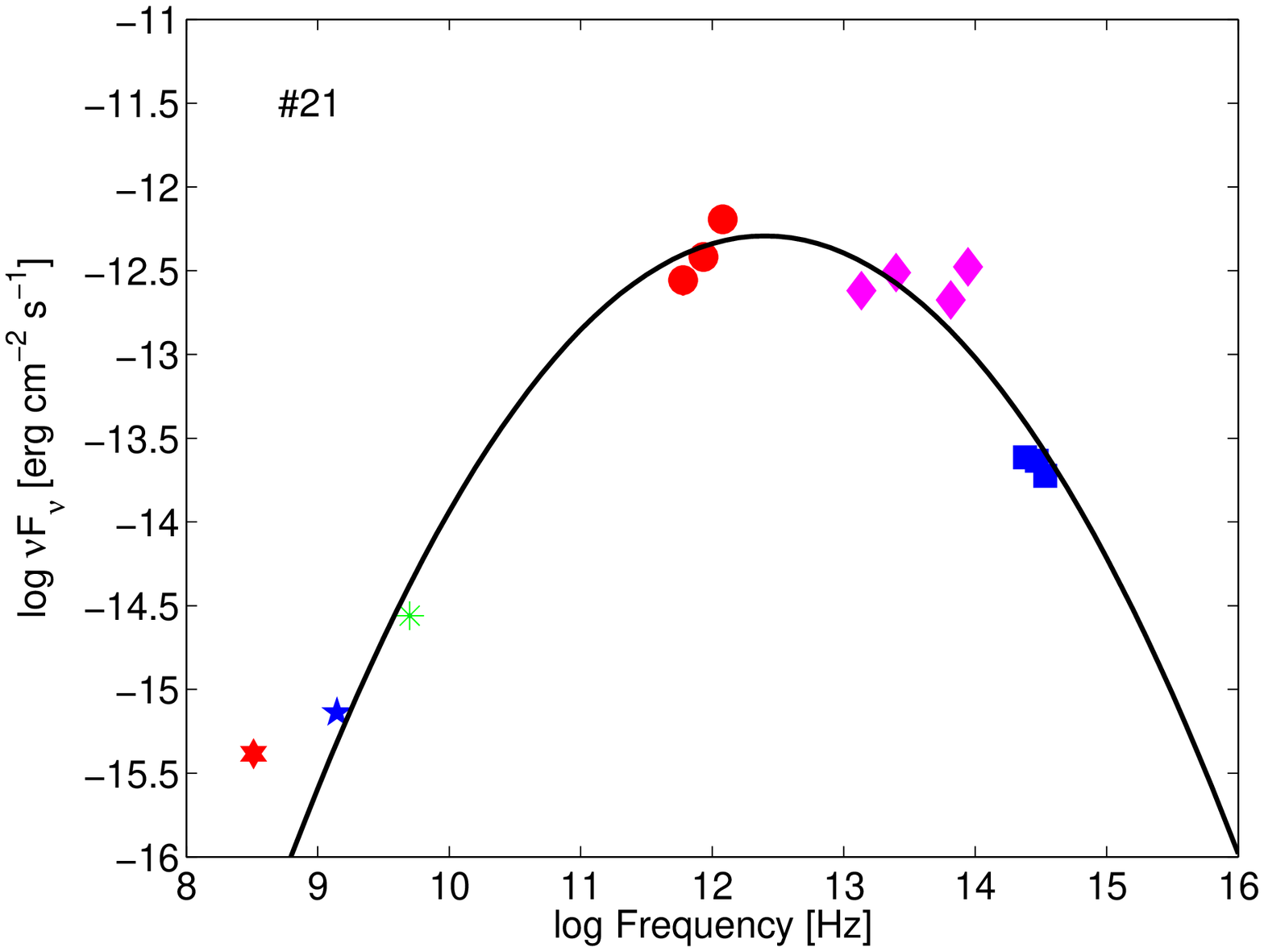}
\end{center}
\caption{SEDs of the 12 candidate blazars in the H-ATLAS equatorial fields (Table~\protect\ref{tab:tabla_candidates}). The meaning of data points and of the solid and dashed lines is the same as in Fig.~\protect\ref{fig:sed_blazars}. The dot-dashed lines are host galaxy templates: a Seyfert 2 galaxy from the SWIRE library \citep{polletta2007} for objects \#~10, 16, and 17; a dusty galaxy template with $log(L_{\rm dust}/L_\odot)>11.5$ \citep{smith2012}
for objects \#~11 and 18; the Arp~220 template from the SWIRE library for objects \#~12, 14, and 19. As for Fig.~\protect\ref{fig:sed_blazars}, these templates are shown for illustration only: no fit of the data was attempted. }\label{fig:sed_candidates}
\end{figure*}

\section{Spectral Energy Distributions} \label{sec:SEDs}

Four out of the 9 known blazars in the H-ATLAS fields have counterparts in the 2-yr {\it Fermi}-LAT catalogue \citep[][see Table~\protect\ref{tab:tabla_confirm8}]{2FGL}. The blazar J090910.1+012135 happened to be observed twice by {\it Herschel}, on 2009 Nov 22 and 2010 May 30, and showed a strong variability, thus providing a direct confirmation of the non-thermal, jet-related nature of its sub-mm emission. Its 250, 350 and $500\,\mu$m flux densities increased from 160, 194 and $266\,$mJy, respectively, at the first epoch, to 258, 338 and $425\,$mJy at the second epoch. This source was also found to be variable also in $\gamma$-rays: its {\it Fermi}-LAT variability index is 159.5.

Figure~\ref{fig:sed_blazars} shows the spectral energy distributions (SEDs) from radio to UV frequencies of the 9 known blazars in the H-ATLAS equatorial fields. To the data from the NASA Extragalactic Database (NED) and to the H-ATLAS data we have added the near-IR photometry provided by the VISTA Kilo-degree INfrared Galaxy survey \citep[VIKING; ][]{Sutherland2012,Fleuren2012} and by the WISE survey\footnote{http://irsa.ipac.caltech.edu/Missions/wise.html}  \citep{Wright2010}.  In addition, 5 of our catalogued blazars and 9 of our blazar candidates were detected by the Giant Metrewave Radio Telescope (GMRT) survey of H-ATLAS equatorial fields at 325 MHz (Mauch et al., in preparation). The measured flux densities are given in Table~\ref{tab:GMRT} and shown in Figs.~\ref{fig:sed_blazars} and \ref{fig:sed_candidates}. The 0.325--1.4\,GHz (FIRST) spectral indices of 4 catalogued blazars are flat or inverted ($\alpha_{0.325}^{1.4} > -0.5$, $S_\nu \propto \nu^\alpha$) while all the 9 blazar candidates have $\alpha_{0.325}^{1.4} < -0.5$. Three of the latter objects (\#\,10, 15, and 18) have 1.4 GHz NVSS flux densities larger than the FIRST ones by a factor $>2$. Since FIRST has a much higher angular resolution than NVSS, this suggests that these sources are extended. As noted in Section\,\ref{sec:Medicina}, however, a steep low-frequency spectrum is an indication, but is not necessarily a proof, that the object is not a blazar since the low-frequency emission may come from a component different from the relativistic jet. In addition, all these data are far from simultaneous. Hence any fit should be dealt with great caution and can only be considered as purely indicative.

\begin{table}
\begin{center}
\begin{tabular}{cc | cc}
\hline ID & $S_{325\rm MHz} $  [mJy] & ID & $S_{325\rm MHz}$  [mJy] \\
\hline
 1  &  500 $\pm$ 15 & 12 &  255 $\pm$ 13 \\
 2 &  157 $\pm$ 5  & 15  &  155 $\pm$ 5 \\
 3 &  986 $\pm$ 31  & 16 &  156 $\pm$ 6 \\
 4 &  4354 $\pm$ 124  &  17  &  72 $\pm$ 5 \\
 8 &  815 $\pm$ 26  & 18  &  210 $\pm$ 7 \\
10  &  602 $\pm$ 21 & 19  &  84 $\pm$ 15 \\
11  &  283 $\pm$ 9  & 21  &  127 $\pm$ 12 \\
\hline
\end{tabular}
\caption{Flux densities at 325 MHz and their errors obtained with the Giant Metrewave Radio Telescope for 5 confirmed blazar and 9 new candidates.}\label{tab:GMRT}
\end{center}
\end{table}

As usual \citep[e.g.,][]{fossati98}, we fit the synchrotron peak with a second order polynomial (i.e., a parabola). The estimated rest-frame synchrotron peak frequencies (in terms of $\nu L_\nu$) are in the range $12.8 \le \log (\nu_{\rm peak, Hz}) \le 14.2$, with a median $\log \nu_{\rm peak, median, Hz}\simeq 13.4$. These objects are thus mostly LSPs, but with an extension in the ISP region. In only one case (\#~3, see \cite{gnuevo10}) do we have enough high-energy data to allow a meaningful estimate of the shape of the inverse Compton peak (not shown in the corresponding panel of Fig.~\ref{fig:sed_blazars}). For this object, the Compton dominance, defined as the ratio between the inverse-Compton and synchrotron peak luminosities is $\nu_{p,IC}L\nu_{p,IC}/\nu_{p,S}L\nu_{p,S} \simeq 8.6 $. Although 3 other blazars have $\gamma$-ray measurements, they are missing X-ray data and therefore the rising part of the inverse Compton peak is essentially unconstrained.

Blazar \#~3 also shows indications of a UV excess that can be attributed to emission from the accretion disk \citep{DermerSchlickeiser1993,GhiselliniTavecchio2009}, as already pointed out by \cite{gnuevo10}. Some hints of a UV bump can also be discerned on the SEDs of blazars \#~4, 6, 7, and 9.  Under the standard assumption that the accretion disk emission is a combination of black-bodies with temperatures depending on the distance from the central black-hole [eq.~(1) of \cite{GhiselliniTavecchio2009}] the black hole mass, $M_{\rm bh}$, can be estimated as
\begin{equation}
{M_{\rm bh}\over 10^9\,M_\odot} \simeq 1.2\left({\eta\over 0.1}\right)^{-{1 \over 2}}\left({T\over 2\times 10^4\,\hbox{K}}\right)^{-2}\left({L_{\rm d}\over 10^{45}\,\hbox{erg}/\hbox{s}}\right)^{-{1 \over 2}}
\end{equation}
where $\eta$ is the mass to light conversion efficiency, $T$ is the maximum black-body temperature, and $L_{\rm d}$ is the total luminosity of the accretion disk. The lack of sufficient simultaneous data hampers reliable estimates of the key quantities $T$ and $L_{\rm d}$, hence of $M_{\rm bh}$ and of the accretion rate $\dot{M_{\rm bh}}=L_{\rm d}/(\eta c^2)$. Tentative values are $M_{\rm bh}\sim \hbox{few}\times 10^9\,M_\odot$, $\dot{M_{\rm bh}}\sim 0.2\hbox{--} 1\,M_\odot\,\hbox{yr}^{-1}$, and Eddington ratios $L_{\rm d}/L_{\rm Edd}\sim 0.2\hbox{--}1\times 10^{-2}$.

The {\it Herschel}/SPIRE colours of blazars \#~1 and 6 seem to indicate a contamination by a star-forming host galaxy with $L_{\rm FIR}\sim 10^{12}\,L_\odot$, i.e. a star-formation rate $\hbox{SFR}\simeq 100$--$170\,M_\odot\,\hbox{yr}^{-1}$ \citep{Kennicutt1998,Lapi2011}. The association of a blazar with a star-forming galaxy is at odds with the notion that blazar hosts are passive ellipticals, as generally found for radio, X-ray and $\gamma$-ray selected blazars \citep{Kotilainen1998,ODowdUrry2005,Kotilainen2007,LeonTavares2011,Giommi2011}.  To check the indication that a significant fraction of sub-mm selected blazars may indeed be associated with star-forming galaxies we have inspected the photometric data available in the NED for 14 catalogued blazars with counterparts in the {\it Planck} ERCSC at both 545 and 857 GHz (see Section\,\ref{sec:blazar_phot}). We have excluded BZUJ1325-4301, i.e. Centaurus A, because of the difficulty of removing the contribution of the extended emission and BZBJ1136+1601 because its classification as a blazar is dubious. For 5 (i.e. $\simeq 40$ percent)  of these objects (BZUJ0840+1312, BZQJ0921+6215, BZQJ1559+0304, BZQJ1719+0817, and BZUJ2219-4710) the  sub-mm continuum appears to be dominated by thermal dust emission. Somewhat surprisingly, 3 of the 5 {\it Planck} ERCSC objects (those labelled BZQ) and both H-ATLAS blazars with possibly/likely star-forming hosts are classified in BZCAT as flat-spectrum radio quasars, although quasar-type blazars are expected to outshine the host galaxy.

As for our blazar candidates, although the available, generally non-simultaneous, data are insufficient to draw firm conclusions, especially on account of the strong blazar variability, we may note that 4 of them  (\#~13, 15, 20, and 21) seem to have SEDs approximately described by a second order polynomial (see Fig.~\ref{fig:sed_candidates}), the usual representation of the synchrotron peak, and thus consistent with being blazar SEDs. The SEDs of the other candidates seem to be strongly contaminated, or dominated, by emission from a dusty star-forming galaxy. Extensive follow-up observations are necessary to assess the nature of these sources. Multi-frequency radio observations, already planned, can establish whether or not our candidates have the ``flat'' high-frequency spectrum typical of blazars. Variability and polarization are further indicators of the blazar nature.

\section{Number counts}\label{sec:counts}

The H-ATLAS survey allows us to put the first meaningful constraints on $500\,\mu$m blazar counts, thus providing a test on evolutionary models and, in consequence, on the underlying physics. For example, the model by \cite{dezotti05} assumes, for all blazars, a flat radio spectral index followed by a parabolic decline above a synchrotron peak frequency that increases with decreasing radio luminosity according to the `blazar sequence' model \citep{fossati98}. On the other hand \cite{tucci11} find that the data favour a broad range of peak frequencies, so broad that any trend with luminosity is blurred. They also argue for different distributions of break frequencies, $\nu_m$, for BL Lacs and FSRQs, in the sense that the former objects have substantially higher values of $\nu_m$, implying that their synchrotron emission comes from more compact regions. Both points are supported by the study of \citet{Giommi2011}.

The \cite{dezotti05}  model predicts $0.21$ blazars$/\hbox{deg}^2$ brighter than 35 mJy at $500\,\mu$m, i.e. $\simeq 28$ blazars over the H-ATLAS Phase 1 area.  At the same flux density limit the \cite{tucci11} predictions are in the range 0.068--0.12 blazars$/\hbox{deg}^2$, implying about 9--16 blazars over the Phase 1 area, the lower values referring to the C2Ex model, the higher to the C2Co model.

As discussed in Section\,\ref{sec:candidates}, in the H-ATLAS equatorial fields totalling $\simeq 135\,\hbox{deg}^2$ we have found 9 catalogued 
blazars and 12 blazar candidates. Two of the catalogued blazars show signs of a substantial host galaxy contribution to the $500\,\mu$m flux density, so that the pure non-thermal emission could well be below our flux density threshold. As for candidate blazars,  there are indications that the $500\,\mu$m flux density of most (8) of them is dominated or at least heavily contaminated by thermal dust emission (see Section\,\ref{sec:SEDs}).

The contamination in the optical/near-IR translates into anomalous colours and may thus introduce incompleteness in the blazar selection based on diagnostics of the kind discussed in Section\,\ref{sec:blazar_phot}. Moreover, the combination of thermal dust and synchrotron emission may result in very red sub-mm colours, so red that the H-ATLAS source extraction procedure, which starts from the identification of $>2.5\sigma$ peaks at ${250\mu{\rm m}}$, may miss objects brighter than our $500\,\mu$m threshold. One such object was indeed recovered by looking for signals in the $500\,\mu$m maps at the positions of known blazars. \cite{rigby11} estimate an incompleteness of 12 percent for $>5\sigma$ $500\,\mu$m sources. The incompleteness may be higher at the $4\sigma$ limit adopted here and for particularly red sub-mm sources. On the other hand, any additional contribution to the observed flux densities biases high the counts compared to the case of pure synchrotron, envisaged by the models.

The present analysis suggests that the second effect (excess $500\,\mu$m flux density due to the contribution of dust in the host galaxy) is the dominant one. If only 4 of our blazar candidates have the $500\,\mu$m flux density  dominated by the non-thermal emission, adding them to the 7 catalogued blazars with probably uncontaminated $500\,\mu$m flux density, we end up with $\sim 11$ blazars brighter than $S_{500\mu{\rm m}}=35\,$mJy in our area, in good agreement with the predictions of \cite{tucci11} C2Ex model, which is also favoured by other data. In any case, the $500\,\mu$m blazar counts at our flux density limit are substantially below the predictions of the \cite{dezotti05} model. This adds to the growing evidence \citep[e.g.][]{statprop} that the synchrotron peak frequency of most blazars occurs at lower frequencies than envisaged by the `blazar sequence' scenario.

\section{Conclusions}\label{sec:conclusions}
The sub-mm selection emphasizes blazar flavours that are marginal or missing in the familiar radio or X-ray selected samples. Two out of the 9 known blazars in the H-ATLAS fields (\#~1 and \#~6) and 8 of the 12 blazar candidates with $S_{500\mu\rm m}\ge 35\,$mJy detected by the H-ATLAS survey in its equatorial fields  show evidence of a dust emission peak at sub-mm wavelengths. 
While we must caution that, as noted in Section \,\ref{sec:blazar_phot}, the H-ATLAS identification of the blazar \#~1 needs to be further checked, and that the nature of candidate blazars is still to be established and most of them might not be blazars at all, the indication that some blazar hosts are endowed with active star formation is supported by the fact that the sub-mm continuum of 5 of the 12 catalogued {\it Planck} ERCSC  blazars detected at both 545 and 857 GHz appear to be dominated by thermal dust emission. This shakes the notion that blazar hosts are passive ellipticals.  Also, somewhat surprisingly, several of these sub-mm selected blazars with possibly/likely star-forming hosts are classified as flat-spectrum radio quasars, although quasar-type blazars are expected to outshine the host galaxy.

The thermal dust contamination of the sub-mm emission complicates the comparison of the observed blazar counts with model predictions. On one side the additional contribution to the observed flux densities leads to an overestimate of the counts compared to the case of pure non-thermal emission, to which models refer. On the other side, the contamination distorts the colours so that the blazars may be missed by diagnostics of the kind discussed in Section\,\ref{sec:blazar_phot}. The present investigation however suggests that the surface density of blazars brighter than 35 mJy at $500\,\mu$m is lower than predicted by the \cite{dezotti05} model, based on the blazar sequence scenario that envisages an anti-correlation between the synchrotron peak frequencies and the radio luminosity. On the other hand, such surface density is consistent with the predictions by \cite{tucci11} who adopted a broad distribution of synchrotron peak frequencies at all luminosities, which would have the effect of lowering the effective synchrotron peak frequency for the bright sources of interest here.

The sub-mm selection may also bring to light blazars with other peculiar properties. In at least one case (blazar \#~6, classified as a flat-spectrum radio quasar) there is evidence of the coexistence of the far-IR/sub-mm bump attributable to the host galaxy with the UV bump interpreted as the thermal emission from the accretion disk. Again this is at odds with the notion that the detection of the UV bump should generally be alternative to the detection of the emission from the host galaxy. In the case of FSRQ the latter is outshone by the AGN emission, while in the case of BL Lacs the weak (or absent) line emission is indicative of a faint thermal emission from the disk. Thus, the accretion disk emission should be detectable only for FSRQs and the host galaxy should be visible only for BL Lacs.

The rest-frame synchrotron peak frequencies (in terms of $\nu L_\nu$) of the 9 catalogued H-ATLAS blazars are in the range $12.8 \le \log \nu_{\rm peak, Hz} \le 14.2$, with a median $\log \nu_{\rm peak, median, Hz}\simeq 13.4$. These objects are thus mostly LSPs, but with an extension in the ISP region.

At this stage all conclusions must be considered as only tentative because of the poor statistics. This will improve as soon as data for the full H-ATLAS survey, covering an area 4 times larger, will be available. The sub-mm selected blazar sample will be further augmented by other {\it Herschel} surveys such as the Herschel Multi-tiered Extragalactic Survey \citep[HerMES,][]{hermes} and the Herschel Virgo Cluster Survey \citep[HeVICS,][]{hevics}. A complementary view of sub-mm selected blazars is being provided by the {\it Planck} surveys, which cover the whole sky with detection limits more than an order of magnitude brighter than achieved by the H-ATLAS survey at similar wavelengths.

\section{Acknowledgements}
The authors acknowledge partial financial support from the Spanish Ministerio de Econom\'ia y Competitividad projects AYA2010-21490-C02-01, AYA2010-21766-C03-01 and CSD2010-00064, from ASI/INAF Agreement I/072/09/0 for the
Planck LFI activity of Phase E2, from MIUR PRIN 2009, and from INAF through the PRIN 2009 ``New light on
the early Universe with sub-mm spectroscopy''. MLC thanks the Spanish Ministerio de Econom\'ia y Competitividad for a Juan de la Cierva fellowship. JGN thanks the Spanish CSIC for a JAE DOC fellowship.

This research has made use of the NASA/IPAC Extragalactic Database (NED), of SDSS DR8 data and of data products from the Wide-field Infrared Survey Explorer (WISE).

The NED is operated by the Jet Propulsion Laboratory, California Institute of Technology, under contract with the National Aeronautics and Space Administration. WISE is a joint project of the University of California, Los Angeles, and the Jet Propulsion Laboratory/California Institute of Technology, funded by the National Aeronautics and Space Administration.

Funding for SDSS-III has been provided by the Alfred P. Sloan Foundation, the Participating Institutions, the National Science Foundation, and the U.S. Department of Energy Office of Science. The SDSS-III web site is http://www.sdss3.org/. SDSS-III is managed by the Astrophysical Research Consortium for the Participating Institutions of the SDSS-III Collaboration including the University of Arizona, the Brazilian Participation Group, Brookhaven National Laboratory, University of Cambridge, University of Florida, the French Participation Group, the German Participation Group, the Instituto de Astrof\'isica de Canarias, the Michigan State/Notre Dame/JINA Participation Group, Johns Hopkins University, Lawrence Berkeley National Laboratory, Max Planck Institute for Astrophysics, New Mexico State University, New York University, Ohio State University, Pennsylvania State University, University of Portsmouth, Princeton University, the Spanish Participation Group, University of Tokyo, University of Utah, Vanderbilt University, University of Virginia, University of Washington, and Yale University.

The Herschel-ATLAS is a project with Herschel, which is an ESA space observatory with science instruments provided by European-led Principal Investigator consortia and with important participation from NASA. The H-ATLAS website is http://www.h-atlas.org/.

\begin{landscape}
\begin{table}
\centering
\begin{tabular}{|c|l|r|r|r|r|r|r|c|c|c|c}

\hline ID & H-ATLAS IAU ID  & RA       &      DEC    &    $S_{250\mu\rm m}$  &     $ S_{350\mu\rm m}$  &   $ S_{500\mu\rm m}$  &   $ S_{\rm FIRST}$  & Distance & z   & BZCAT NAME & {\it Fermi}  2FGL NAME \\
            &   &   $[^{\circ}]$ &    $[^{\circ}]$   &     [mJy]  &     [mJy] &     [mJy] &     [mJy] & [arcsec]& & & \\
\hline
\hline     1  & HATLAS J083949.3+010436 & 129.95576   &      1.07669    &   50.50   &    50.68  &    42.83  &    443.71  & 9.72  & 1.123 & $BZQJ0839+0104$ &   2FGL J0839.6+0059 \\
\hline     2   &  HATLAS J090940.3+020000 & 137.41804   &      2.00013    &   40.29  &    64.49 &    73.87  &    305.76  & 8.67  &   - & $BZBJ0909+0200$ &    2FGL J0909.6+0158\\
\hline     3   & HATLAS J090910.1+012135  &137.29245   &      1.35986   &  258.11   &  337.74  &  424.86  &    559.64 & 1.42  &1.024 & $BZQJ0909+0121$ &    2FGL J0909.1+0121\\
\hline    4   & HATLAS J115043.8-002355  & 177.68276   &    -0.39885   &    37.90   &    63.28  &    79.22 &  2803.17  &  1.90  &1.976 & $BZQJ1150-0023$ & \\
\hline    5   & HATLAS J113245.7+003427 & 173.19054   &      0.57434   &    65.81   &    73.66  &    57.25  &    468.98  &  1.43  &   - & $BZBJ1132+0034$ &  2FGL J1132.9+0033  \\
\hline    6   &  HATLAS J113302.9+001545 & 173.26229   &      0.26256   &    59.41   &    48.56  &    40.35  &    214.96  &  3.99 & 1.173 & $BZQJ1133+0015$ &  \\
\hline    7   & HATLAS J113320.1+004054  & 173.33408   &      0.68185   &    35.61   &     43.24 &     46.90 &     312.49 & 2.40  &1.633 & $BZQJ1133+0040$ &  \\
\hline   8   &  HATLAS J141004.6+020306 & 212.51958   &     2.05174   &  114.23   &  152.90   &  178.05  &    291.62  &  0.80  &   - & $BZBJ1410+0203$ &  \\
\hline   9   &  HATLAS J140412.1-001325 & 211.05040   &   -0.22360   &   18.00    &   32.00       & 45.00     &   516.32  & 0.08  &1.217  &  $BZQJ1404-0013$ &  \\
\hline
\end{tabular}
\caption{BZCAT blazars found in the H-ATLAS 9\,h, 12\,h and 15\,h fields.  Objects  BZBJ0909+0200 and  BZQJ0909+0121 are those found by Gonz\'alez-Nuevo et al. (2010) in the H-ATLAS Science Demonstration Phase field. Objects $\#1$--8  are listed in H-ATLAS catalogues while object $\#9$ lies near a border of the map and was excluded when making the catalogues. The distance between the H-ATLAS source and the FIRST low radio frequency counterpart is shown. The redshifts are from the BZCAT.}
\label{tab:tabla_confirm8}
\end{table}

\begin{table}
\centering
\begin{tabular}{|c|l|r|r|r|r|r|r|c|r|r|c}
\hline   ID &  H-ATLAS IAU ID & RA     &         DEC   &    $S_{250\mu\rm m}$  &     $ S_{350\mu\rm m}$  &   $ S_{500\mu\rm m}$  &   $ S_{\rm FIRST}$  &  Distance  & FIRST ID  &  FIRST RA & FIRST DEC  \\
          &     &   $[^{\circ}]$ &    $[^{\circ}]$   &     [mJy]  &     [mJy] &     [mJy] &     [mJy] & [arcsec]&  &[hms] &[dms] \\
\hline
\hline  % 19  &
10  &  HATLAS J143405.2-002307 & 218.52182  &   -0.38549  &   73.40    &   70.02  &  41.95   &    59.92   & 3.01 &     J143405.2-002310  &  14 34 05.248  &  -00 23 10.76 \\ % & 1 \\
\hline  % 20  &
11  & HATLAS J144119.7+002511  & 220.33240  &    0.41993  &    52.44    &   60.79  &  41.15   &    57.32  & 7.31 &     J144119.4+002506  &  14 41 19.412  &  +00 25 06.88 \\ %  &  1 \\
\hline % 21  &
  12 & HATLAS J140931.9-010054  & 212.38320  &   -1.01520  &   43.92    &   38.33  &  35.31   &    63.61   & 7.60 &     J140931.4-010053  &  14 09 31.465  &  -01 00 53.83 \\%   &   1,2 \\
\hline % 22  &
    13 & HATLAS J144930.2+023636 & 222.37620  &    2.61022  &    24.43    &   43.89  &  46.27   &    50.09   & 4.32 &     J144930.3+023632  &  14 49 30.310  &  +02 36 32.48 \\ % &  0 \\
\hline  % 23  &
  14 & HATLAS J113622.0+004854  & 174.09173  &     0.81505  &   99.44    &   65.30  &  43.97   &  151.74  & 2.40 &     J113622.0+004851  &  11 36 22.041  &  +00 48 51.80 \\ % &  1 \\
\hline % 24  &
    15  &  HATLAS J085418.9-003612 & 133.57879  &  -0.60352  &    56.07    &   68.31  &  48.88   &    25.43    & 7.24 &     J085418.6-003619    &   08 54 18.690   & -00 36 19.14 \\ % & 0\\
\hline % 25  &
    16 &  HATLAS J114616.8-001937  & 176.57016  &  -0.32712  &    95.44    &   69.55  &  37.28   &    22.17     & 1.11 &     J114616.8-001936    &   11 46 16.820   & -00 19 36.53 \\ % &  0\\
\hline  % 26  &
   17 & HATLAS J114203.3+005133  & 175.51381  &    0.85938  &  147.77    &   66.52  &  36.79   &    15.92     & 3.00 &     J114203.4+005136   &   11 42 03.440   &   +00 51 36.05 \\ % & 0 \\
\hline  % 27  &
   18 & HATLAS J142631.1+014226  & 216.62988  &    1.70600  &    30.70    &   43.20  &  49.80   &    23.39     & 5.92 &     J142631.4+014231   &   14 26 31.400   & +01 42 31.25  \\ % & 1\\
\hline  % 28  &
   19 & HATLAS J143743.4-005237  & 219.43098  &  -0.87711  &    53.80    &   49.90  &  47.70   &     35.40   & 8.76 &     J143742.8-005240    &   14 37 42.880   & -00 52 40.64  \\ % & 0 \\
\hline  % 29  &
   20 &  HATLAS J145146.2+010610 & 222.94281  &    1.10293  &    35.60    &   60.30  &  62.70   &     41.78    &  9.83 &     J145145.9+010619   &   14 51 45.950   & +01 06 19.10 \\ % &  0\\
\hline % 30  &
    21 & HATLAS J144937.4+004135 & 222.40611  &    0.69322  &   53.42  &   44.550  &  46.26  &    51.93    & 8.19 &    J144937.9+004131  &  14 49 37.946  &  +00 41   31.66   \\ % & 2 \\ %steep

\hline
\end{tabular}
\caption{New blazar candidates found in the H-ATLAS 9\,h, 12\,h and 15\,h fields. Redshifts are not available for any of these objects.}
\label{tab:tabla_candidates}
\end{table}
\end{landscape}

\begin{table*}
\begin{tabular}{|r|r|r|r|r|r|c|c|c|c}
\hline    RA     &         DEC  &    $ S_{545\rm GHz}$  &  $S_{600\rm GHz}$  & $ S_{857\rm GHz}$  &   $ S_{\rm FIRST}$  & $z$   &   BZCAT NAME  & 2FGL NAME  \\
        $[^{\circ}]$ &    $[^{\circ}]$   &     [mJy]  &     [mJy] &     [mJy] &     [mJy] & & & & \\
\hline
\hline   
140.2420   &     44.6956  &    1142.91  &      1152.93 &      1190.84 &       1017.00 &  2.190   &   BZQJ0920+4441 & 2FGLJ0920.9+4441 \\
\hline   
133.7000   &     20.1131  &    3618.51  &      3377.79 &      2616.81 &       1512.00 &  0.306    &   BZBJ0854+2006 &  \\
\hline   
343.4920   &     16.1481  &   20585.74 &    20086.98 &    18340.72 &    12657.00 &  0.859    &   BZQJ2253+1608 &  \\
\hline   
187.2750   &      2.0458   &     6937.50 &      6193.01 &     4065.28  &    54991.00 &  0.158   &   BZQJ1229+0203 &  \\
\hline   
194.0420   &     -5.7894  &     4701.58 &      4370.46 &     3333.65  &      9711.00 &  0.536    &   BZQJ1256-0547 &  \\
\hline   
60.9710     &   -36.0869  &     2338.46 &      2134.30 &     1521.04  &      1151.00 &  1.417     &   BZQJ0403-3605  &  \\
\hline   
84.7120     &   -44.0853  &     4065.40 &      3826.54 &     3057.01  &      3729.00 &  0.892      &   BZBJ0538-4405  &  \\
\hline   
140.3960   &     62.2628  &     1234.73 &      1495.28 &     3041.21  &        946.00 &  1.446    &   BZQJ0921+6215 &  \\
\hline  
130.1979   &     13.2066  &     2801.51 &      3263.22 &     5745.42  &      2614.00 &  0.681     &   BZUJ0840+1312 &  2FGLJ0840.7+1310 \\
\hline
\end{tabular}
\caption{BZCAT blazars with a {\it Planck}-ERCSC counterpart at both 545 and 857 GHz (see text). The 600\,GHz flux densities were computed interpolating between measurements at 545 and 857\,GHz.  The redshifts are from the BZCAT.}
\label{tab:tabla_ercsc}
\end{table*}

\end{document}